\documentclass[fleqn,10pt]{wlscirep}
\usepackage[utf8]{inputenc}
\usepackage[T1]{fontenc}
\usepackage{subfigure}
\usepackage{caption}
\usepackage{float}
\usepackage{multirow}
\usepackage{lineno}
\usepackage{ragged2e}
\usepackage{xcolor}
\definecolor{fullred}{rgb}{1 0 0}
\usepackage{bm}

\begin{document}

\title{Plasticity Encoding and Mapping during Elementary Loading for Accelerated Mechanical Properties Prediction}

\author[1,*]{Mathieu Calvat}
\author[1]{Chris Bean}
\author[1]{Dhruv Anjaria}
\author[2]{Haoren Wang}
\author[2]{Kenneth Vecchio}
\author[1,*]{J.C. Stinville}

\affil[1]{Materials Science and Engineering Department, University of Illinois Urbana-Champaign, Urbana, Illinois, USA}
\affil[2]{Department of NanoEngineering, University of California, San Diego, La Jolla, CA, USA}
\affil[*]{e-mail: mcalvat@illinois.edu; jcstinv@illinois.edu}

\begin{abstract}
    Encoding metal plasticity captured from high-resolution digital image correlation (DIC) can be leveraged to predict a wide range of monotonic and cyclic macroscopic properties of metallic materials. To capture the spatial heterogeneity of plasticity that develops in metals, latent space features describing plasticity of a small region are spatially mapped across a large field of view while maintaining the same spatial relationships as the experimental measurements. Latent space feature maps capture the complexity and heterogeneity of metal plasticity as a low-dimensional representation. These feature maps are then used to train a convolutional neural network-based model to predict monotonic and cyclic macroscopic properties. The approach is demonstrated on a large set of face-centered cubic metals, enabling rapid and accurate property prediction. The effects of hyperparameters and training strategies are analyzed, and the extension of the proposed approach to a broader range of metallic materials and loading conditions is discussed.
\end{abstract}

\flushbottom
\maketitle

\thispagestyle{empty}

%\linenumbers

\section*{Keywords}

\justify Plasticity localization -- High Resolution Digital Image Correlation -- Encoding -- Properties prediction -- Metallic materials and alloys

\section{Introduction}

\justify With the advent of accelerated measurement techniques in materials science, it is now possible to collect complex datasets that provide 2D or 3D representations of a material's structure and local behavior \cite{annurev:/content/journals/10.1146/annurev-matsci-080921-102621, GIANOLA2023101090, KAUFMANN2020178, CHENG2025101429,Miracle2021131}. This advancement goes beyond the traditional reliance on single metrics (or a limited set of metrics) to describe material microstructure or behavior. Instead, large field-of-view spatial measurements can be utilized to capture the hierarchical structure and associated variability within materials \cite{annurev:/content/journals/10.1146/annurev-matsci-080921-102621,WINIARSKI2021113315}. Similarly, it is now possible to rapidly capture the local behavior of materials through full-field, high-resolution measurements \cite{annurev:/content/journals/10.1146/annurev-matsci-080921-102621,Black2023}. Measurements of deformation can be performed over large fields of view at high resolution, enabling the observation of material deformation at small scales with remarkable detail and to be statistically representative \cite{annurev:/content/journals/10.1146/annurev-matsci-080921-102621,Black2023}.

\justify For instance, in the field of metals, it is possible to rapidly characterize microstructure at the nanometer scale across large fields of view in both 2D \cite{WINIARSKI2021113315} and 3D \cite{Echlin2021, Stinville2022}. Moreover, these extensive datasets can include local behavioral information. Examples include acquiring large multi-modal 2D and 3D microstructure datasets through point diffraction measurements or full-field deformation and damage assessments \cite{Burnett2019, Stinville2022}. These techniques are becoming increasingly accessible and automated \cite{KAUFMANN2024101192}, creating opportunities to autonomously and rapidly identify processing-structure-property relationships, thereby accelerating materials design.

\justify However, achieving this goal presents significant challenges. Relating these complex datasets, such as maps of microstructure and deformation, to material macroscopic properties requires innovative solutions. The complexity arises because these multi-modal datasets can consist of billions of data points, making it difficult to establish direct connections to macroscopic material properties, even with advanced data-driven analysis tools. 

%\justify In the specific context of the mechanical behavior of metals, it is now possible to acquire local measurements of nanometer scale plasticity over large fields of view, enabling detailed observation of local deformation behavior. For example, X-ray techniques can provide detailed information about elastic fields at the microstructure scale over extensive areas. Similarly, automated electron microscopy allows for the mapping of surface plasticity across unprecedented fields of view in a relatively short amount of time. 

\justify The correlation of spatial measurement of microstructure and deformation allows one to understand the intrinsic effects of microstructural features on plasticity \cite{Burnett2014,CHARPAGNE2021117037, STINVILLE2022111891, CHARPAGNE2020110245}. This understanding ultimately guides the design of new materials and optimized microstructures. However, these approaches are very time consuming and require the use of complex analysis tools to draw relationships with the different modalities of such correlated datasets. Another promising approach is to leverage recent advancements in machine learning to directly predict macroscopic properties from spatial fields describing plasticity \cite{doi:10.1073/pnas.1922210117}. Plasticity, and the way it evolves (plastic flow), is a critical factor in controlling the mechanical behavior of metals \cite{Stinville2022}. At small scales, plasticity develops in a highly inhomogeneous manner, leading to localized plasticity (plastic localization) that dominates mechanical performance \cite{Stinville2022}.

\justify Here, we propose shifting the focus from linking plasticity to microstructure to directly establishing the relationship between plasticity and macroscopic properties. Ultimately, microstructure governs plasticity, which, in turn, controls mechanical behavior. Therefore, plasticity localization full-field measurements contain the necessary information to predict mechanical properties directly from measurements of plasticity. In this study, a Variational AutoEncoder (VAE) approach has been considered to construct low-dimensional representation to describe the spatial distribution and characteristics of metal plasticity. These latent space representations are subsequently used to predict the macroscopic mechanical properties. In this study, a large set of face-centered cubic (FCC) alloys has been considered, including wrought and additively manufactured materials to train and validate our approach. Plasticity during early stages of monotonic loading is captured and encoded to predict a range of macroscopic properties. These include those related to monotonic loading, such as yield strength, ultimate tensile strength, elongation, and hardening parameters as well as fatigue properties. Finally, the localization of plasticity and its use to predict mechanical properties, the structure of the latent space associated with plasticity and the application of this approach for accelerating material design are then discussed. A specific methodology is then proposed to efficiently extend such a database.

\begin{figure}[htbp]
    \centering
    \includegraphics[width=1\textwidth]{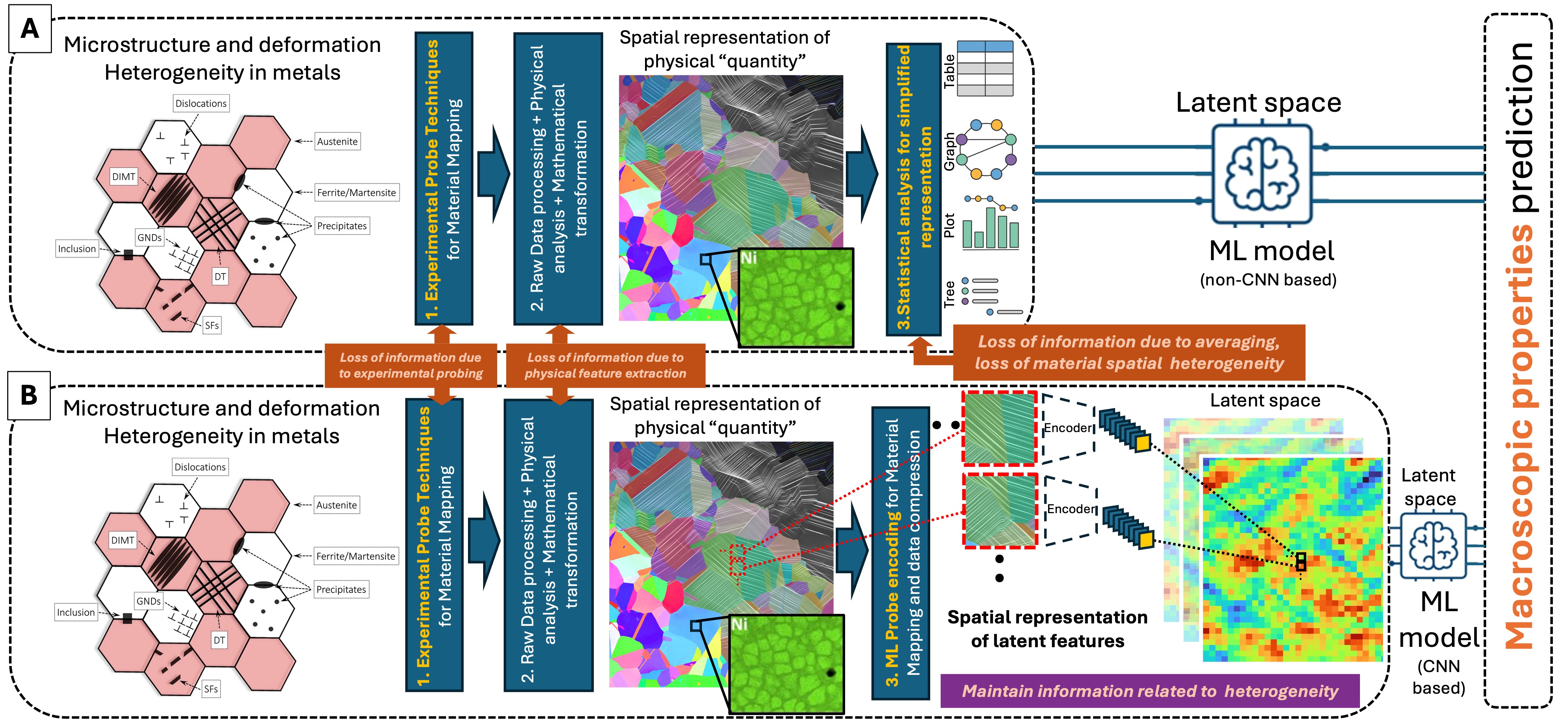}
    \caption{\textbf{Comparison between the conventional method and the proposed method.} \textbf{(A)} In the conventional approach, microstructure and deformation fields are measured using experimental probing techniques. Subsequently, discrete metrics or simplified representations are extracted to relate these measurements to macroscopic properties. Significant information loss occurs due to data averaging, notably the loss of spatial heterogeneity, a critical factor influencing macroscopic properties. Adapted from \cite{Raabe2020,Stinville2022,GRIFFITHS2021110815} with permission from Springer and Elsevier. \textbf{(B)} The proposed approach differs primarily in the final step, where encoders compress the data into a latent space. This compressed representation is combined with latent space feature mapping to preserve spatial heterogeneity, while remaining sufficiently reduced for effective use in data-driven models.}
    \label{fig:Intro}
\end{figure}

\justify As illustrated schematically in Fig. \ref{fig:Intro}, one conventional approach (Fig. \ref{fig:Intro}(A)) for predicting mechanical properties using data-driven analysis relies on extracting simplified metrics from experimentally correlated measurements. These metrics typically include grain and precipitate size, shape, texture, or simplified representations such as graph networks \cite{Thomas2023}. Such methods, however, only capture averaged information and consequently lose critical details related to microstructural or deformation heterogeneity. In contrast, we propose the method detailed in Fig. \ref{fig:Intro}(B), which employs encoding (data compression) and latent space feature mapping to simplify the measured fields while preserving all essential information and more importantly spatial heterogeneity.

\section{Database and Method}

\subsection*{An FCC materials database} \label{sec:database}

\justify A total of nine FCC materials, including both wrought and additively manufactured specimens, were investigated in this study. These materials are listed in Table \ref{tab:materials} in the methods section. In the material denominations, \textbf{W}, \textbf{AM}, \textbf{RX}, and \textbf{AB} refer to wrought, additively manufactured, recrystallized, and as-built conditions, respectively. The materials studied include Invar, Nickel 600, Steel 330, three variants of the nickel-based superalloy Inconel 718 (In718) processed through different routes, a copper-nickel-tin alloy (Cu77NiSn), a cobalt-based alloy (Co76A), and a stainless steel 316L (Steel316). All materials exhibit an FCC matrix with strengthening mechanisms that include solid-solution and/or precipitate strengthening. Their chemical compositions are provided in Table \ref{tab:materials} in the methods section and their associated monotonic and cyclic properties are reported in Table \ref{tab:props} in the methods section. Under the investigated deformation conditions, all materials deform primarily via dislocation slip.

\subsection*{High-resolution digital image correlation data} \label{sec:data}

\justify High-resolution digital image correlation (HR-DIC) was performed using the Heaviside technique (H-DIC) \cite{Valle2017}, which accounts for discontinuity in the subset during computations. This approach, outlined in detail in reference\cite{Valle2017}, allows quantitative capture of irreversible deformation mechanisms associated with plastic deformation of metallic materials (here, slip in the present study) \cite{Bourdin2018}. This technique achieves a resolution of a few nanometers in displacement induced by deformation. More details concerning the parameters used in this study can be found in the methods section. 

\begin{figure}[htbp]
    \centering
    \includegraphics[width=1\textwidth]{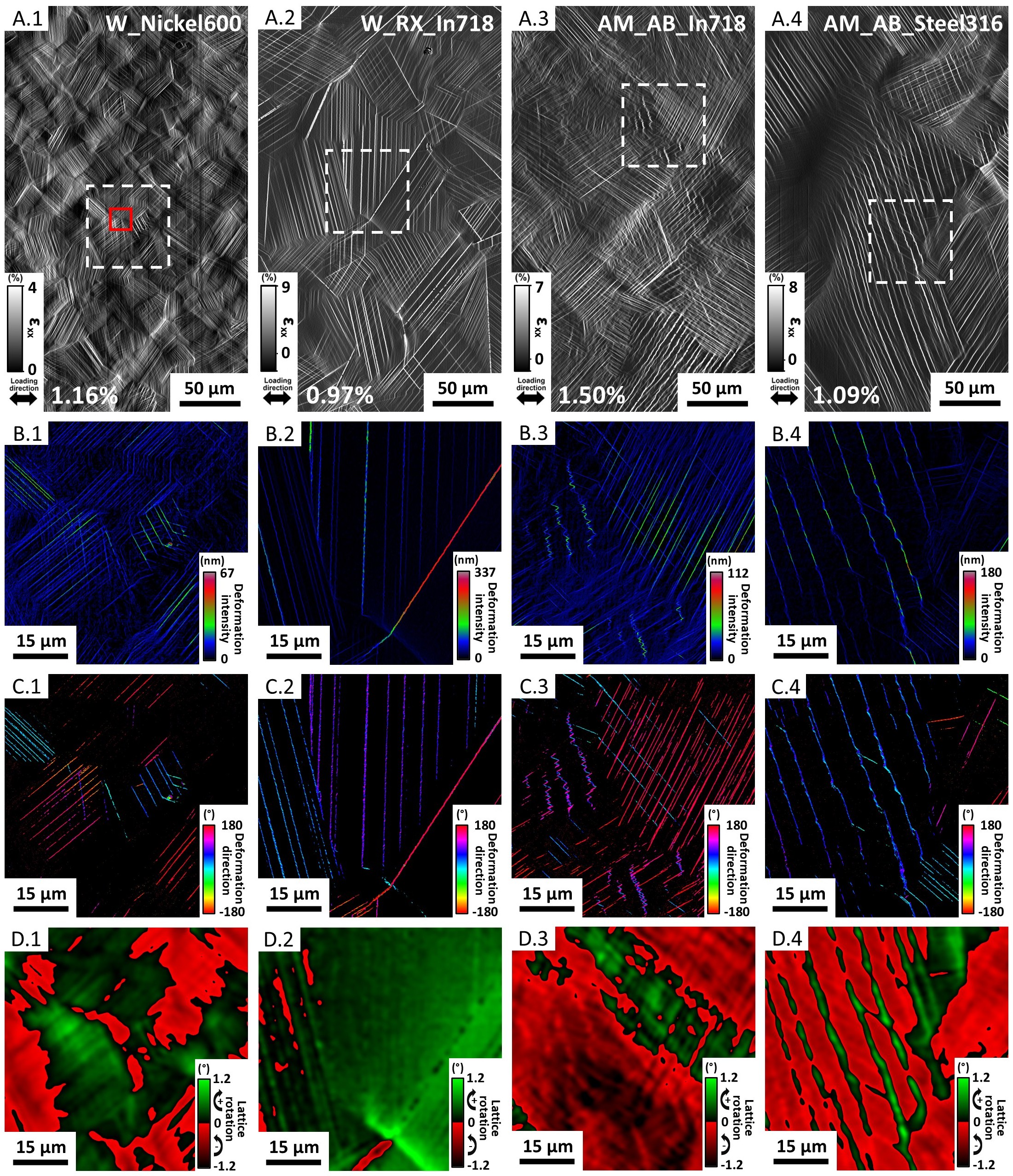}
    \caption{\textbf{Spatial distribution, morphologies and characteristics of plastic deformation events, a dataset of metal plasticity}. \textbf{(A)} Longitudinal  strain $\varepsilon_{xx}$, \textbf{(B)} plastic deformation intensity, \textbf{(C)} plastic deformation direction and \textbf{(D)} lattice rotation for a \textbf{(.1)} wrought Nickel600 (W\_Nickel600), \textbf{(.2)} wrought and recrystallized Inconel 718 (W\_RX\_In718), \textbf{(.3)} additively manufactured Inconel 718 in its as-built condition (AM\_AB\_In718) and \textbf{(.4)} an additively manufactured 316L in its as-built condition (AM\_AB\_Steel316) at macroscopic plastic strains of 1.16\%, 0.97\%, 1.50\% and 1.09\%, respectively. All materials were deformed at room temperature and the tensile direction is horizontal. As an example, a 256 $\times$ 256 window with a red border is shown in \textbf{(A.1)}.} 
    \label{fig:hrdic}
\end{figure}

\justify To illustrate the information that can be extracted from HR-DIC, small regions of the longitudinal strain $\varepsilon_{xx}$ (strain along the loading direction) maps are shown in Fig. \ref{fig:hrdic}(A.1-A.4) for a selection of materials used in this study (wrought Nickel600, wrought Inconel 718, additively manufactured Inconel 718 in its as-built structure and additively manufactured stainless steel 316L in its as-built structure, respectively) during monotonic tensile loading. Bands of concentrated strain are observed, indicating discrete plastic deformation events (i.e. slip) that develop during the deformation of metallic materials. With this approach, the intensity of each individual plastic deformation event can be quantified in nanometers and corresponds to the in-plane displacement amplitude associated with the local shearing induced along the deformation event \cite{Bourdin2018,Texier2024,Anjaria2024}. Examples of such intensity maps are given in Fig. \ref{fig:hrdic}(B.1 - B.4) for reduced regions of interest, represented by white boxes in Fig. \ref{fig:hrdic}(A). For each plastic deformation event associated with an in-plane displacement, a shearing direction is also defined and measured by HR-DIC. Examples are shown in Fig. \ref{fig:hrdic}(C.1-C.4) using the angle between the shearing direction and the trace of the deformation event. Additionally, metal deformation induces non-localized rotation between localized deformation events that can be measured with HR-DIC \cite{STINVILLE2020110600}. These are shown in Fig. \ref{fig:hrdic}(D.1-D.4) using a divergent color scale, with clock wise rotation in green. For a more comprehensive visualization of the dataset, all the longitudinal strain components, $\varepsilon_{xx}$, are displayed in Multimedia Component 1.

\justify All these maps are overlaid at pixel-level resolution, forming a comprehensive multi-modal dataset that describes alloy deformation under a given loading condition. Significant variations in deformation characteristics are observed between alloys, with some exhibiting straight bands indicative of planar slip, while others display more complex morphologies. The density, intensity, length, and geometric orientation of these bands differ substantially from alloy to alloy. In addition, the rotation field induced between deformation events varies between different alloys.

% The plastic deformation intensity and direction correspond to the in-plane slip amplitude (in nanometers) and in-plane direction (in degrees), respectively

% Both plastic deformation direction and the lattice rotation are represented using a colormap with discontinuity at $0^{\circ}$.

\justify Each alloy was deformed monotonically at room temperature ato four different macroscopic plastic strains, referred to as $\varepsilon_{xx}^p$, in the range of 0.13\% to 2.43\%. In addition, one alloy, the Invar, was also deformed at an elevated temperature of 500$^\circ$C. The result is a large set of HR-DIC maps, each composed by 1.1 million pixels. As is, the data can not be used for training an ML-based model to predict the mechanical properties due to the size of the maps but also the limited number of individual maps. Therefore, an approach has been developed to both extract and compress data while mitigating the loss of information. 

\begin{figure}[htbp]
    \centering
    \includegraphics[width=1\textwidth]{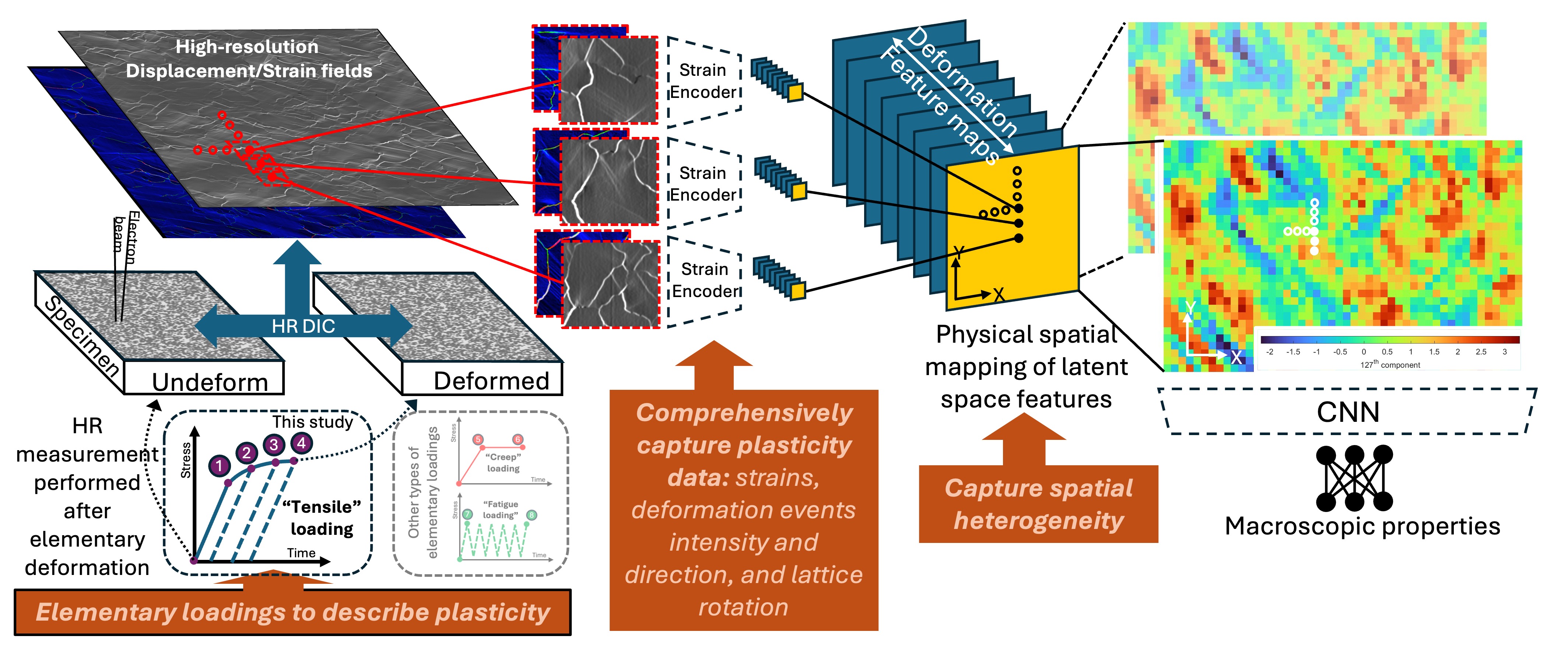}
    \caption{\textbf{Encoding of plasticity and spatial mapping of latent space features for mechanical properties prediction.} The proposed approach consists of performing full-field and high-resolution measurements of deformation over a large field of view during elementary (simple) loading conditions. The generated deformation maps are divided into smaller regions, each of which is encoded using a trained variational autoencoder. Subsequently, the latent space features obtained for each small region are mapped according to their spatial locations. Latent space feature maps are generated and used to predict mechanical properties.}
    \label{fig:principle}
\end{figure}

\justify After acquiring the data, the approach described schematically in Fig. \ref{fig:principle} is applied to predict the mechanical properties. This approach involves dividing each of the acquired HR-DIC maps into smaller regions, each of which is encoded using a trained VAE. Subsequently, the latent space features extracted from each small region are spatially mapped according to their positions on the HR-DIC maps. Finally, the generated latent space feature maps are utilized to predict mechanical properties through a convolutional neural network (CNN) architecture. Prior to applying this method, encoders must be trained individually for each HR-DIC modality, and the CNN used for predicting mechanical properties from latent space feature maps must also be trained; these processes are detailed in the following section.

% The detection itself strongly depends on the sensitivity of the considered model and might struggle with low-intensity events \cite{Bean2025}. 

\subsection*{Encoding of HR-DIC full-field measurements} \label{sec:encoding}

\begin{figure}[htbp]
    \centering
    \includegraphics[width=1\textwidth]{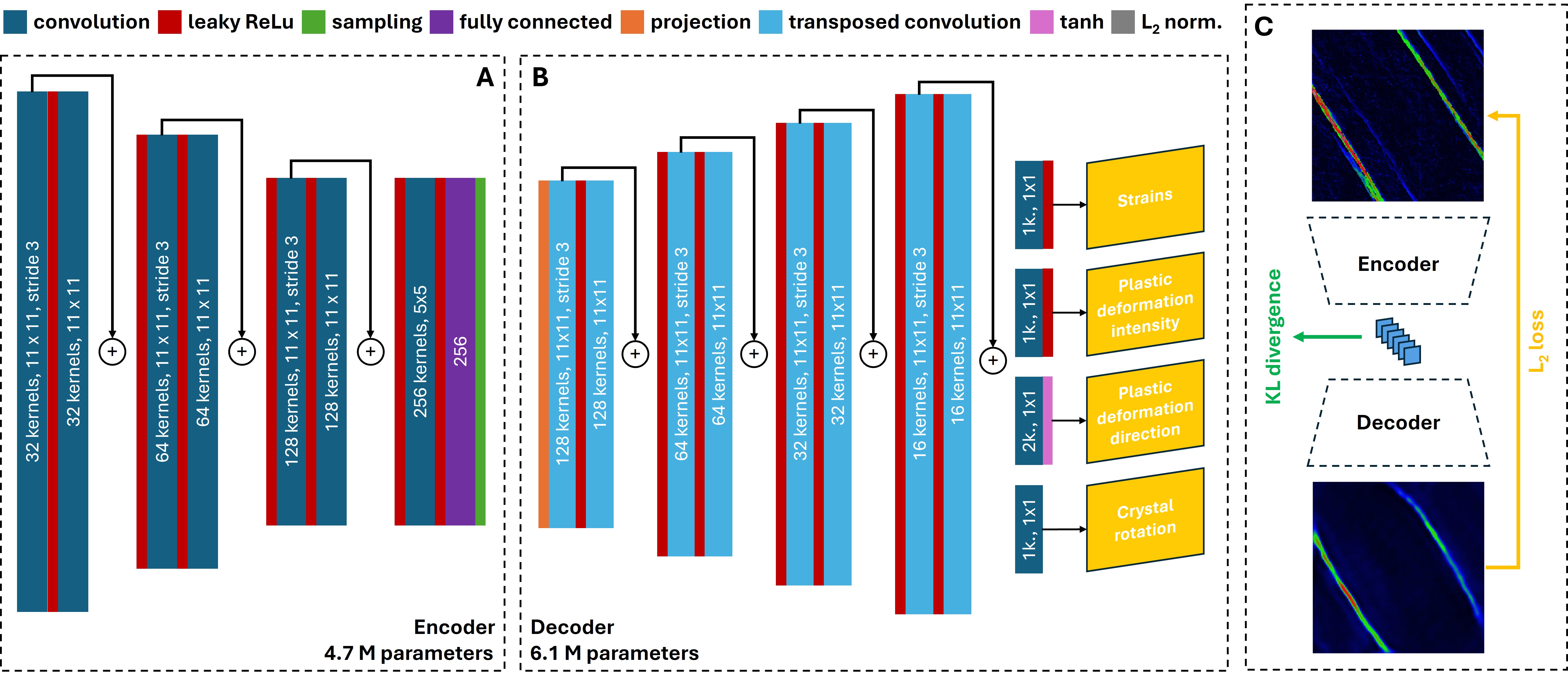}
    \caption{\textbf{(A,B)} Encoders and decoders are employed to accurately represent reduced HR-DIC regions corresponding to various modalities obtained from the HR-DIC measurements. Detailed VAE architecture descriptions for the \textbf{(A)} encoder and \textbf{(B)} decoder. \textbf{(C)} Training methodology of the VAE with an example concerning the plastic deformation intensity.}
    \label{fig:vae}
\end{figure}

\justify A VAE architecture has been developed and trained to encode the different modalities obtained from the HR-DIC measurements (strains, plastic deformation intensity and direction, and lattice rotation) into low-dimensional latent space representations. The developed architecture is given in Fig. \ref{fig:vae}(A,B) and contains two different CNNs, the \textit{encoder} used to project the given HR-DIC data into a 128-dimension latent space, and the \textit{decoder} to restore the original data from a low-dimensional representation. Both of these CNNs use large convolution kernels to avoid pixelation and to facilitate the detection of straight events. Shortcut connections popularized by ResNets \cite{He2016} have also been used on all convolution layers without downscaling or upscaling operations. No normalization layer has been considered in these CNN architectures due to the physical meaning of the HR-DIC data. The \textit{encoder} structure remains the same across the considered modalities of the HR-DIC data and only the final layers of the \textit{decoder} have been modified. To solve the issue of rotation continuity associated with plastic deformation direction, these maps have been transformed to polar coordinates and the last leaky ReLU layer was replaced by a hyperbolic tangent activation.
 
\justify The training principle of the VAE architecture is illustrated in Fig. \ref{fig:vae}(C) and integrates a pixel-to-pixel $L_2$ loss (estimated between the original image and its restored counterpart) and a Kullback-Leibler divergence term to ensure that the learned distributions converge to a standard normal distribution \cite{Doersch2016}. Unlike conventional images, HR-DIC data are not bounded to a given range (i.e. between 0 to 255, for instance). For this reason, the individual $L_2$ loss has been inversely scaled by the maximum value in each window. This allows reducing the focus of the VAE on the most intense deformation events. Additional details about the training procedure are given in the methods section.

\begin{figure}[htbp]
    \centering
    \includegraphics[width=1\textwidth]{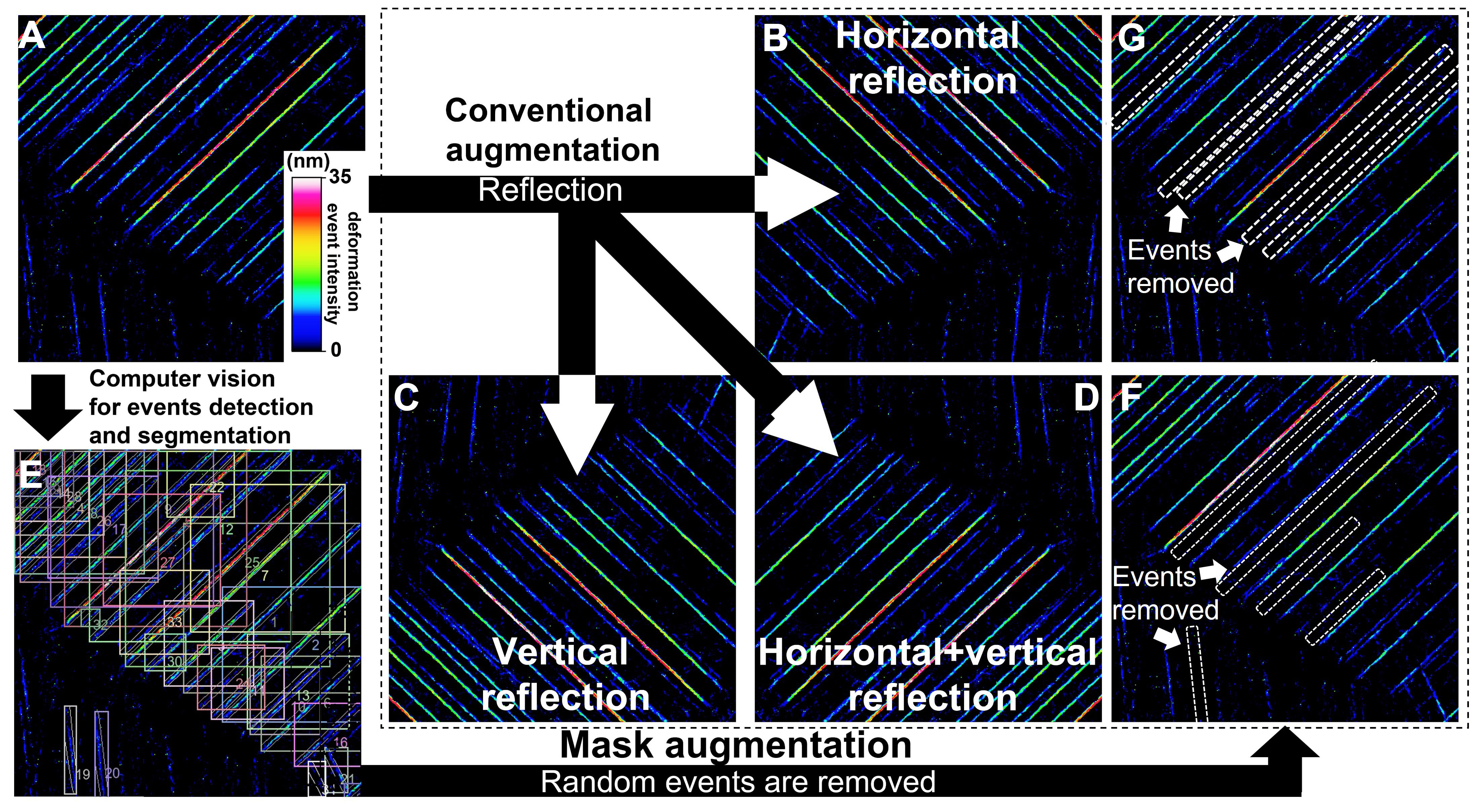}
    \caption{\textbf{(A)} Deformation event intensity map for a selected region of interest in an Inconel 718 alloy after plastic deformation of 0.97\%. \textbf{(B,C,D)} The first set of augmentations applied, consisting of horizontal, vertical, and combined horizontal-vertical reflections. \textbf{(E)} The computer vision framework developed in \cite{Bean2025} enables the detection and segmentation of individual deformation events within the HR-DIC data. \textbf{(F,G)} Events detected from the CV framework are randomly removed from the initial maps to generate additional realistic HR-DIC maps.}
    \label{fig:aug}
\end{figure}

\justify Several types of augmentations have been considered in this study as they allow an increased amount of data seen by the VAE architecture and consequently improves the reconstruction quality. As the chosen augmentations must maintain a realistic plasticity localization field, crop and rescale, Gaussian blur, pixel value alteration are not suitable operations for data augmentations. For instance, Gaussian blur may reduce the description of sharp events by adapting the latent space to also reconstruct blurry events. In addition to horizontal, vertical, and vertical plus horizontal reflections (see Fig. \ref{fig:aug}(B,C and D), respectively), masking augmentations based on computer vision (CV) event detection were used. A previously developed and trained CV model \cite{Bean2025} is used to detect deformation events as shown in Fig. \ref{fig:aug}(E). The identified masks were used to randomly remove plastic deformation events and replace them by a Gaussian distribution generated background. A given region containing several deformation events can result in a large number of augmented versions. These augmentations have been illustrated in Fig. \ref{fig:aug}(F and G) for the modality of plastic deformation intensity and can be applied to the other HR-DIC modalities, with the exception of lattice rotation. For the considered data set, reflections and masking augmentations lead to a data multiplying factor of 4 and about 10\textsuperscript{6}, respectively. For the latter, the augmentation factor is not identical for all reduced regions, and depends directly on the number of bands found in this region.

\begin{figure}[htbp]
    \centering
    \includegraphics[width=1\textwidth]{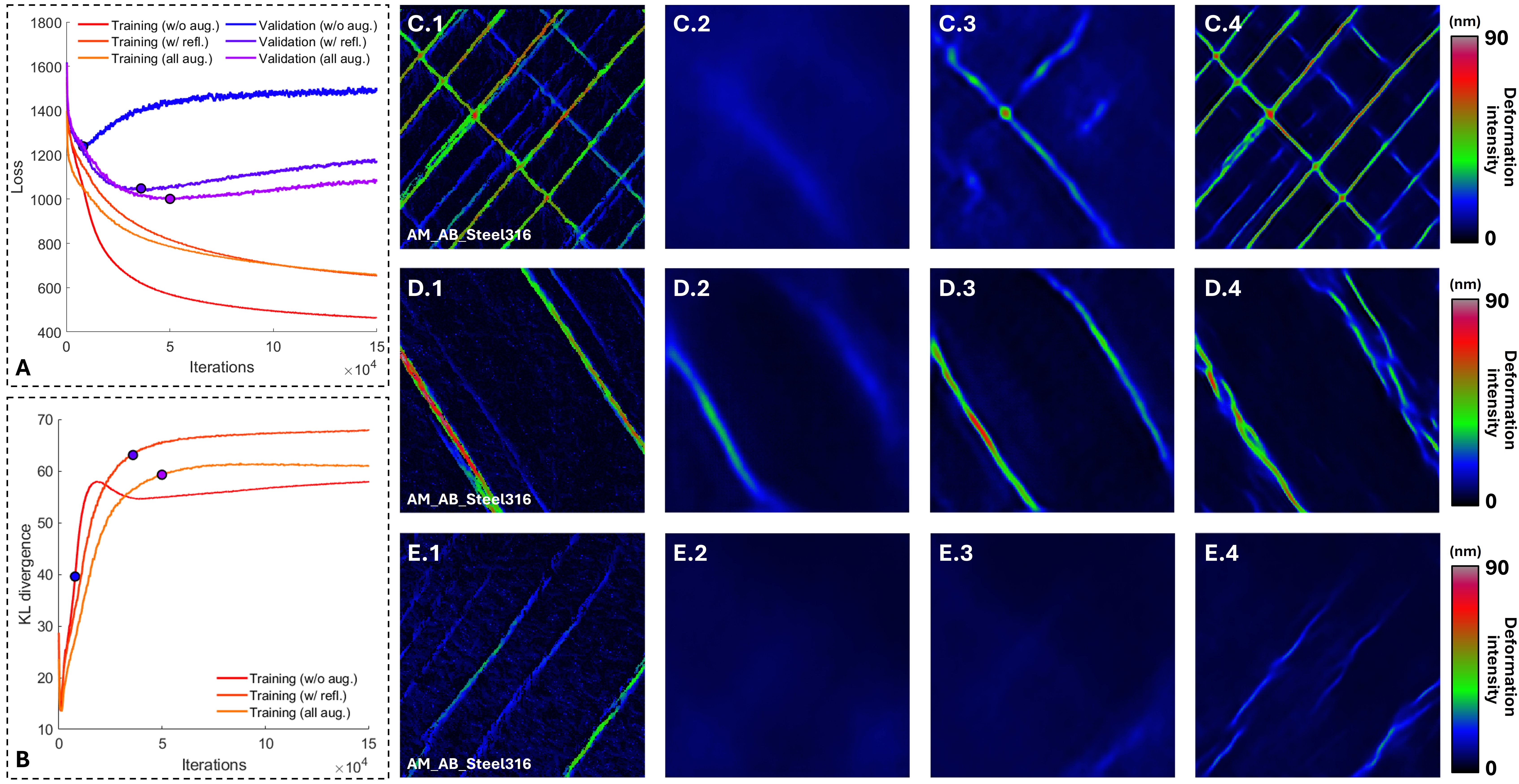}
    \caption{Smoothed \textbf{(A)} scaled $L_2$ loss and \textbf{(B)} KL divergence associated to the trainings with different augmentations. Plastic deformation intensity windows from AM\_AB\_Steel316 at 1.10\% of macroscopic plastic strain and selected in \textbf{(C)} training set, \textbf{(D, E)} validation set. \textbf{(.1)} Original window compared to different stages of training: \textbf{(.2)} early training, \textbf{(.3)} optimal training and \textbf{(.4)} late training. The tensile direction is horizontal.}
    \label{fig:loss}
\end{figure}

\justify Three different trainings were performed to encode/decode the plastic deformation intensity: without augmentation, with reflection augmentations, and with reflections and masking augmentations. The losses corresponding to these trainings are shown in Fig. \ref{fig:loss}(A) for both training and validation sets (without augmentations), smoothed over a 500-iteration window for visualization purposes. For each training, the retained set of weights corresponds to the minimal value associated with the validation set (without augmentations). At first, all losses are similar except for the training loss associated with masking augmentation, since masking randomly hides events and globally reduces the loss. Without augmentation, the training loss decreases faster as the whole dataset is being processed at each epoch. The optimal training is reached after 8000 iterations (160 epochs). As expected, the augmentations allow to achieve a lower minimal validation loss, about 1.15 and 1.2 times lower for reflections and all considered augmentations, respectively. This lower minimum is obtained for a much higher number of epochs: 720 epochs with reflections and 1,000 epochs for all considered augmentations. After this optimal point, while the training loss continues to decrease, the validation loss increases due to over-fitting. Additionally, the KL divergence variation during training is shown in Fig. \ref{fig:loss}(B). At an early stage of training, the KL divergence increases to store more information. The optimal training point is reached during the KL divergence increase, but at a different percentage of the maximum value. During over-fitting, the KL divergence decreases strongly when no augmentation is considered, as it provides a significant leverage on the loss. The divergence associated with the validation set follows the same trend (not shown here).

\justify The evolution of the reconstruction quality during the training without augmentation is shown in Fig. \ref{fig:loss}(C) for the training set and in Fig. \ref{fig:loss}(D,E) for the validation set. The different reconstructions qualities can be compared to the original image given in Fig. \ref{fig:loss}(C.1,D.1,E.1). First, the CNN architecture focuses on reconstructing the background and a blurred version of the image that provides the best leverage on the loss, as shown in Fig. \ref{fig:loss}(C.2) at an early stage of training (halfway to the optimal point). At the optimal point, the training loss is on average lower than the validation loss, resulting in a better reconstruction in Fig. \ref{fig:loss}(C.3) compared to Fig. \ref{fig:loss}(E.3), with some better reconstructions of the validation set (Fig. \ref{fig:loss}(D.3)). During late training (three times past the optimal point), the reconstruction quality of the training set continues to improve between Fig. \ref{fig:loss}(C.3) and Fig. \ref{fig:loss}(C.4). However, reconstruction quality of the validation set deteriorates between Fig. \ref{fig:loss}(D.3) and Fig. \ref{fig:loss}(D.4) due to over-fitting. Specifically, over-fitting mostly affects the reconstruction of close and parallel events, as shown in Fig. \ref{fig:loss}(D.4). Rarely, the quality of the validation set reconstruction improves further, as shown in Fig. \ref{fig:loss}(E.4), but this seems to be limited to windows with very few events of low intensity. 

\begin{figure}[htbp]
    \centering
    \includegraphics[width=1\textwidth]{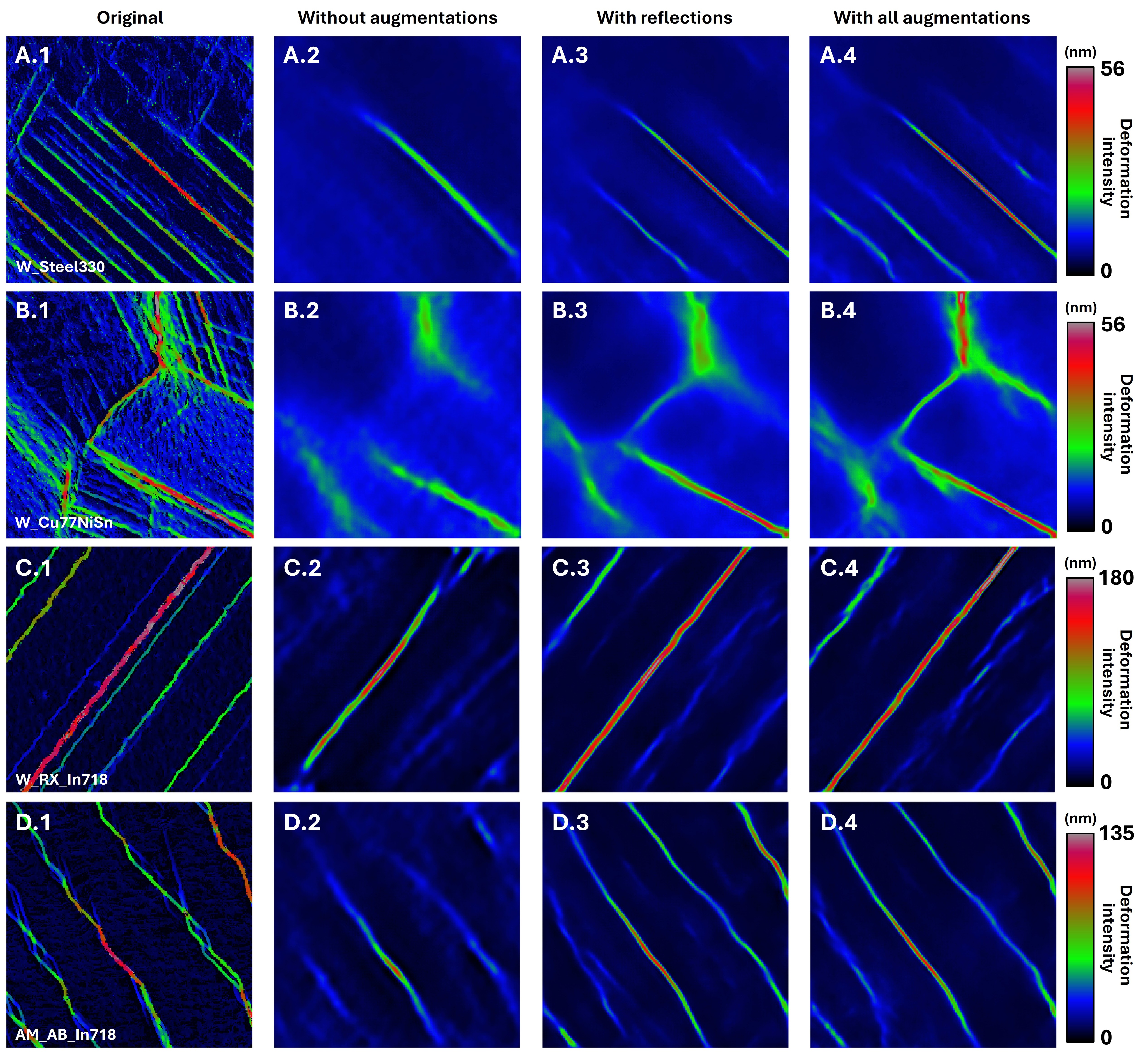}
    \caption{Windows selected within \textbf{(A)} W\_Steel330, \textbf{(B)} W\_Cu77NiSn, \textbf{(C)} W\_RX\_In718 and \textbf{(D)} AM\_AB\_In718, at 0.56\%, 1.02\%, 0.97\% and 1.50\% macroscopic plastic strain, respectively. \textbf{(.1)} Original plastic deformation intensity and reconstructed versions with optimal training (minimum validation loss) and different augmentations: \textbf{(.2)} without augmentations, \textbf{(.3)} with reflections and \textbf{(.4)} with reflections and masking. The tensile direction is horizontal.}
    \label{fig:restoredbln}
\end{figure}

\justify Additionally, Fig. \ref{fig:restoredbln} illustrates the quality of reconstruction of several windows from the wrought Steel 330 (W\_Steel330), the wrought CuNiSn alloy (W\_Cu77NiSn), the wrought Inconel 718 (W\_RX\_In718) and the additively manufactured Inconel 718 in as-built condition (AM\_AB\_In718) at optimal training and for the different augmentations. The original maps are given in Fig. \ref{fig:restoredbln}(A.1,B.1,C.1,D.1). Using reflection augmentations improves the description of the deformation events (number of events and plastic deformation intensity) when comparing Fig. \ref{fig:restoredbln}(A.3,B.3,C.3,D.3) with Fig. \ref{fig:restoredbln}(A.2,B.2,C.2,D.2). The addition of masking augmentations slightly improves the optimal validation loss (Fig. \ref{fig:loss}(A)), inducing an overall improvement in the reconstruction of deformation events. When complex deformation event morphologies are present (see Fig. \ref{fig:restoredbln}(D)), the improvement is not observed to be significant. The VAE trained with both reflections and masking augmentations provides the best validation and will be used to encode HR-DIC data in the following section. 

% Despite the scaled $L_2$ loss used, the most intense deformation events present a better reconstruction as shown in Fig. \ref{fig:restoredbln}() whereas low-intensity events notably found in Fig. \ref{fig:restoredbln}()

\begin{figure}
    \centering
    \includegraphics[width=1\linewidth]{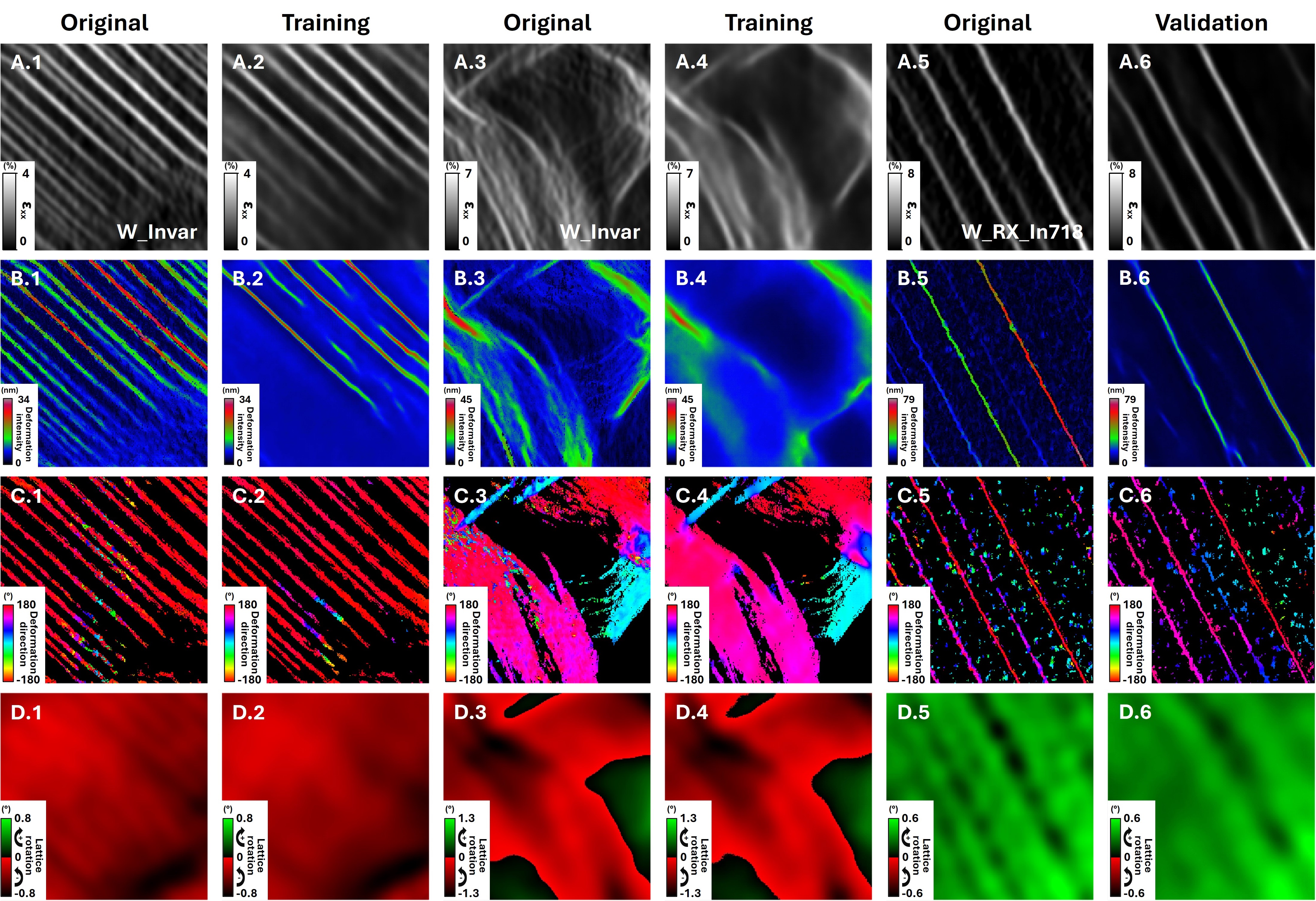}
    \caption{Comparison of the reconstructions of the various HR-DIC data extracted from the same regions: \textbf{(A)} Longitudinal strain, \textbf{(B)} plastic deformation intensity, \textbf{(C)} plastic deformation direction and \textbf{(D)} lattice rotation. The tensile direction is horizontal. \textbf{(.1, .3, .5)} show original windows whereas \textbf{(.2, .4, .6)} illustrate the reconstructions using the different VAE architectures.}
    \label{fig:rec}
\end{figure}

\justify Finally, Fig. \ref{fig:rec} shows the reconstructions of the different HR-DIC data modalities of the same regions at optimal training. Fig. \ref{fig:rec}(A.), (B.), (C.), (D.) display the $\varepsilon_{xx}$ strain, plastic deformation intensity, plastic deformation direction and lattice rotation, respectively. All reconstructions have been produced from networks trained with all the considered augmentations (Fig. \ref{fig:aug}). The proposed approach provides good reconstruction regardless of the modality encoded and decoded. Table \ref{tab:perfo} details the performance of the VAE architectures when reconstructing the plastic deformation intensity of the investigated alloys. The performances have been computed solely on the testing portion of the dataset and are reported in terms of structural similarity \cite{Wang2004,Hore2010} (SSIM) and peak signal-to-noise ratio \cite{Gonzalez2009,Hore2010} (PSNR). The PSNR value has been calculated using the maximum value of the initial HR-DIC modality. The performance referred to as \textbf{1\textsuperscript{st}} is associated with the reconstructions of the HR-DIC data corresponding to the lowest plastic strains given in Table \ref{tab:props} whereas the \textbf{Avg} corresponds to the average performance over the four different steps. Additionally, Table \ref{tab:perfo2} provided in the supplementary materials details the performance for the reconstructions of the other HR-DIC modalities. Performance is high and increases, as previously observed in Fig. \ref{fig:restoredbln}, using augmentations. When comparing the different materials, W\_Invar exhibits the highest SSIM values. In contrast, the AM\_AB\_In718 exhibits the lowest SSIM and PSNR values. A visual inspection of the HR-DIC results reveals that planar slip (characterized by straight bands) dominates in W\_Invar, as observed in most of the investigated materials. However, AM\_AB\_In718 exhibits significant and unusual cross-slip activity (bands are deviating from being straight), making it an exception and less represented within the training set.

\begin{table}[h!]
    \centering
    \setlength{\tabcolsep}{6pt}
    \caption{Performance of reconstruction of the plastic deformation intensity associated to the different trainings.}
    \label{tab:perfo}
    \begin{tabular}{|c||c|c|c|c|c|c|}
        \hline
         & \multicolumn{6}{c|}{Plastic deformation intensity} \\
        Alloy & \multicolumn{2}{c|}{Without augmentations} & \multicolumn{2}{c|}{With reflections} & \multicolumn{2}{c|}{With all augmentations} \\
        Denomination & SSIM & PSNR & SSIM & PSNR & SSIM & PSNR \\
        & 1\textsuperscript{st} -- Avg. & 1\textsuperscript{st} -- Avg. & 1\textsuperscript{st} -- Avg. & 1\textsuperscript{st} -- Avg. & 1\textsuperscript{st} -- Avg. & 1\textsuperscript{st} -- Avg. \\
        \hline
        W\_Invar & 0.594 -- 0.443 & 21.85 -- 21.23 & 0.597 -- 0.453 & 22.23 -- 21.92 & 0.599 -- 0.456 & 22.41 -- 22.08 \\
        \hline
        W\_Nickel600 & 0.404 -- 0.245 & 23.31 -- 20.29 & 0.406 -- 0.249 & 23.56 -- 20.47 & 0.409 -- 0.251 & 23.70 -- 20.58 \\
        \hline
        W\_Steel330 & 0.213 -- 0.133 & 20.20 -- 18.64 & 0.217 -- 0.138 & 20.44 -- 18.87 & 0.219 -- 0.140 & 20.54 -- 18.95 \\
        \hline
        W\_In718 & 0.353 -- 0.247 & 23.21 -- 20.77 & 0.354 -- 0.248 & 23.27 -- 20.90 & 0.354 -- 0.249 & 23.27 -- 20.94 \\
        \hline
        W\_Cu77NiSn & 0.341 -- 0.222 & 22.87 -- 21.03 & 0.345 -- 0.226 & 23.21 -- 21.35 & 0.348 -- 0.229 & 23.37 -- 21.47 \\
        \hline
        W\_RX\_In718 & 0.307 -- 0.229 & 22.65 -- 22.41 & 0.330 -- 0.255 & 23.79 -- 23.50 & 0.334 -- 0.257 & 23.17 -- 23.81 \\
        \hline
        AM\_AB\_In718 & 0.168 -- 0.156 & 20.01 -- 20.79 & 0.176 -- 0.169 & 20.49 -- 21.32 & 0.178 -- 0.174 & 20.51 -- 21.55 \\
        \hline
        AM\_AB\_Co76A & 0.247 -- 0.198 & 21.32 -- 21.46 & 0.253 -- 0.210 & 21.72 -- 22.02 & 0.255 -- 0.213 & 21.77 -- 22.20 \\
        \hline
        AM\_AB\_Steel316 & 0.320 -- 0.219 & 22.01 -- 21.26 & 0.339 -- 0.243 & 22.99 -- 22.33 & 0.345 -- 0.248 & 23.27 -- 22.63 \\
        \hline
    \end{tabular}
\end{table}

\subsection*{Mechanical properties prediction}

\justify The latent space representations of plasticity are utilized to predict mechanical properties. Based on the developed VAE architecture, a single latent space representation (128 latent space features) describes a 17 $\mu$m $\times$ 17 $\mu$m area. The proposed CNN architecture for mechanical properties prediction takes as input $9\times9$ windows (153 $\mu$m $\times$ 153 $\mu$m, i.e. $9\times9$ 128 latent space features) of low-dimensional representations to predict all the macroscopic mechanical properties listed in Table \ref{tab:props} in the methods section. The properties include the yield strength (YS), hardening rate (Q) and exponent (b), ultimate tensile strength (UTS), elongation (El) and fatigue strength (FS). Details on how these macroscopic properties were obtained are provided in the methods section. Rather than using the four macroscopic strain states altogether to predict the macroscopic properties, the model takes as input data from a single deformation state. Recall that each specimen was deformed to four different levels of macroscopic plastic deformation, resulting in four HR-DIC maps for a given modality.

\begin{figure}[htbp]
    \centering
    \includegraphics[width=1\textwidth]{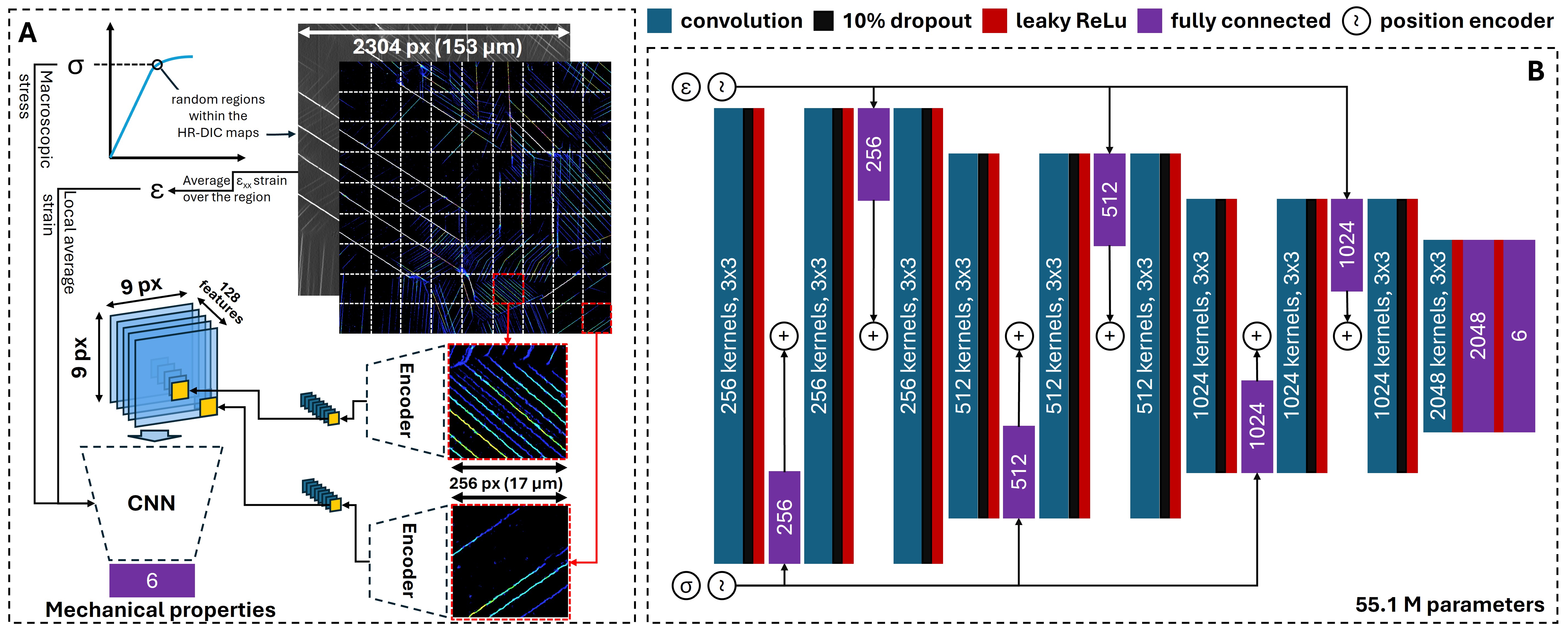}
    \caption{\textbf{(A)} Principle of the CNN used for mechanical properties prediction. Regions of size $9 \times 9$ are extracted from the HR-DIC measurements. Each window is encoded using a VAE, producing 128 latent space feature maps of size $9 \times 9$. These latent space feature maps are used as input to the CNN to predict the six investigated mechanical properties. \textbf{(B)} Detailed architectures of the CNN used for mechanical properties prediction.}
    \label{fig:cnn}
\end{figure}

\justify The predictive approach is illustrated in Fig. \ref{fig:cnn}(A), where, for a given deformation state (i.e. macroscopic plastic deformation), a region of the HR-DIC measurement is selected and encoded. A deformation event intensity map consisting of $9 \times 9$ windows is extracted from the HR-DIC data, with each window individually encoded. This process results in 128 latent space feature maps of size $9 \times 9$, which serve as input to a CNN-based architecture for predicting the six macroscopic mechanical properties considered in this study. Additionally, the local average strain, obtained by averaging the $\varepsilon_{xx}$ maps of the $9 \times 9$ encoded windows, is also used as an input to the CNN architecture. The macroscopic stress, recorded during tensile loading, corresponding to the HR-DIC map is another input for prediction. Stress and strain inputs are utilized through sinusoidal position encoders \cite{Vaswani2017} once normalized (by 5\% and $1500\:\mathrm{MPa}$, respectively). The CNN architecture, detailed in Fig. \ref{fig:cnn}(B), ensures that the local average strain and macroscopic stress are introduced at different depths to prevent the loss of this critical information during convolutions. The spatial resolution is slowly reduced with $3\times3$ convolution kernels while the number of feature maps is increased. No stride has been used to consider potential connectivity of events between adjacent low-dimensional representations. Due to the limited amount of data, 15\% dropout layers were considered.

\begin{figure}[htbp]
    \centering
    \includegraphics[width=1\textwidth]{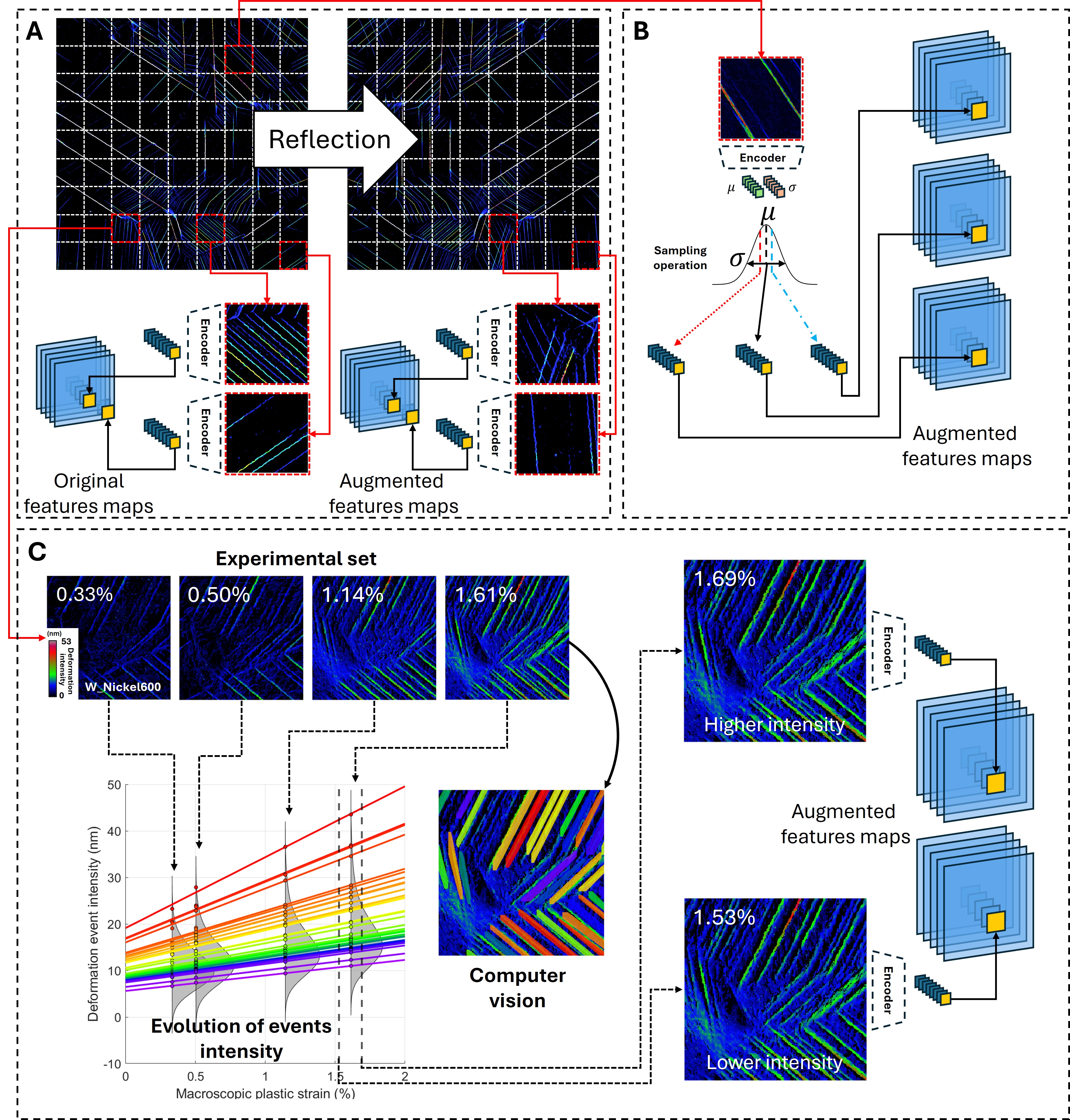}
    \caption{Different augmentations used to train the CNN architecture: \textbf{(A)} conventional horizontal and vertical reflections, \textbf{(B)} augmentation based  on the Gaussian sampling of the VAE architecture and \textbf{(C)} augmentation based on the deformation intensity linear trend as a function of the macroscopic plastic strain.}
    \label{fig:aug2}
\end{figure}

\justify Before becoming operational, the CNN-based model detailed in Fig. \ref{fig:cnn}(B) is trained to simultaneously predict all mechanical properties investigated using an $L_2$ loss function between the predicted and experimentally identified values. All the mechanical properties have been standardized for training, which ensures that all the mechanical properties are within similar ranges thereby improving the training of the CNN architecture. The training dataset comprises 9,216 $9 \times 9$ windows of low-dimensional representations, along with the corresponding macroscopic stress at which the HR-DIC maps were acquired and the average $\varepsilon_{xx}$ strain over the region associated with each $9 \times 9$ window. To mitigate the limited amount of data for training, horizontal, vertical and horizontal plus vertical reflections have first been considered as augmentations as shown in Fig. \ref{fig:aug2}(A). In addition to conventional reflections two specific types of augmentations are considered and are presented in Fig. \ref{fig:aug2}(B and C). The first kind of augmentation, derives naturally from the VAE definition\cite{Kingma2013} and its construction around the VAE sampling layer. The latter takes as input twice the number of dimensions of the latent space, the first half of which are the means $\mu$, while the second half are the standard deviations $\sigma$ of the Gaussian distribution from which the low-dimensional features $z$ are sampled. This sampling operation further introduces variability in the reconstruction, which can be used as an advantage to train the CNN architecture instead of simply training on the means $\mu$ and standard deviations $\sigma$. The final augmentation method is based on the physical considerations of plastic deformation intensity evolution. As a first approximation, a linear relationship exists between macroscopic plastic strain and the intensity of the plastic deformation  events \cite{Bean2025}, which can be leveraged for data augmentation. This approach, illustrated in Fig. \ref{fig:aug2}(C), utilizes HR-DIC data collected during loading. For each window, the evolution of deformation events is measured at four macroscopic stress/strain levels. These events can be identified using a previously developed computer vision approach \cite{Bean2025}, allowing their evolution during loading to be extracted (see Fig. \ref{fig:aug2}(C)). Based on the observed evolution of the relative intensity of each event, new windows can be constructed by interpolating event intensities between the measured deformation steps. Since the intensity of each event follows a predictable trend, as shown in Fig. \ref{fig:aug2}(C), its intensity can be estimated at different macroscopic strains. Subsequently, pixels within the mask obtained from the computer vision approach are scaled by a factor corresponding to the considered strain and evolution of the intensity of events. This process generates new windows, serving as physically realistic augmentations. Strains within $\pm 5\%$ of the experimentally measured values are considered. An important limitation of this approach is that it only modifies the intensity of existing events but does not add or remove events, which occur experimentally but remain limited when only slight changes of macroscopic strain are considered.

\justify For validation of the trained model, HR-DIC measurements were performed on a supplementary sample per material. These additional HR-DIC measurements are only used for validation purpose and are not incorporated into the training set. Both training and validation losses are shown in Fig. \ref{fig:fit}(A.) which have been smoothed over a 10-iteration window for visualization purposes. At each iteration through the training, a validation loss has been computed from 8 regions, without any augmentation or dropout layers. The training was discontinued at 1,000 iterations (8 epochs) as no clear increase was observed for the validation loss. In addition, the relative error is shown in Fig. \ref{fig:fit}(A.2) and has been computed over all the mechanical properties; similarly to the loss, no significant change or increase is observed. 

\section{Results and Discussion}

\begin{figure}[htbp]
    \centering
    \includegraphics[width=1\textwidth]{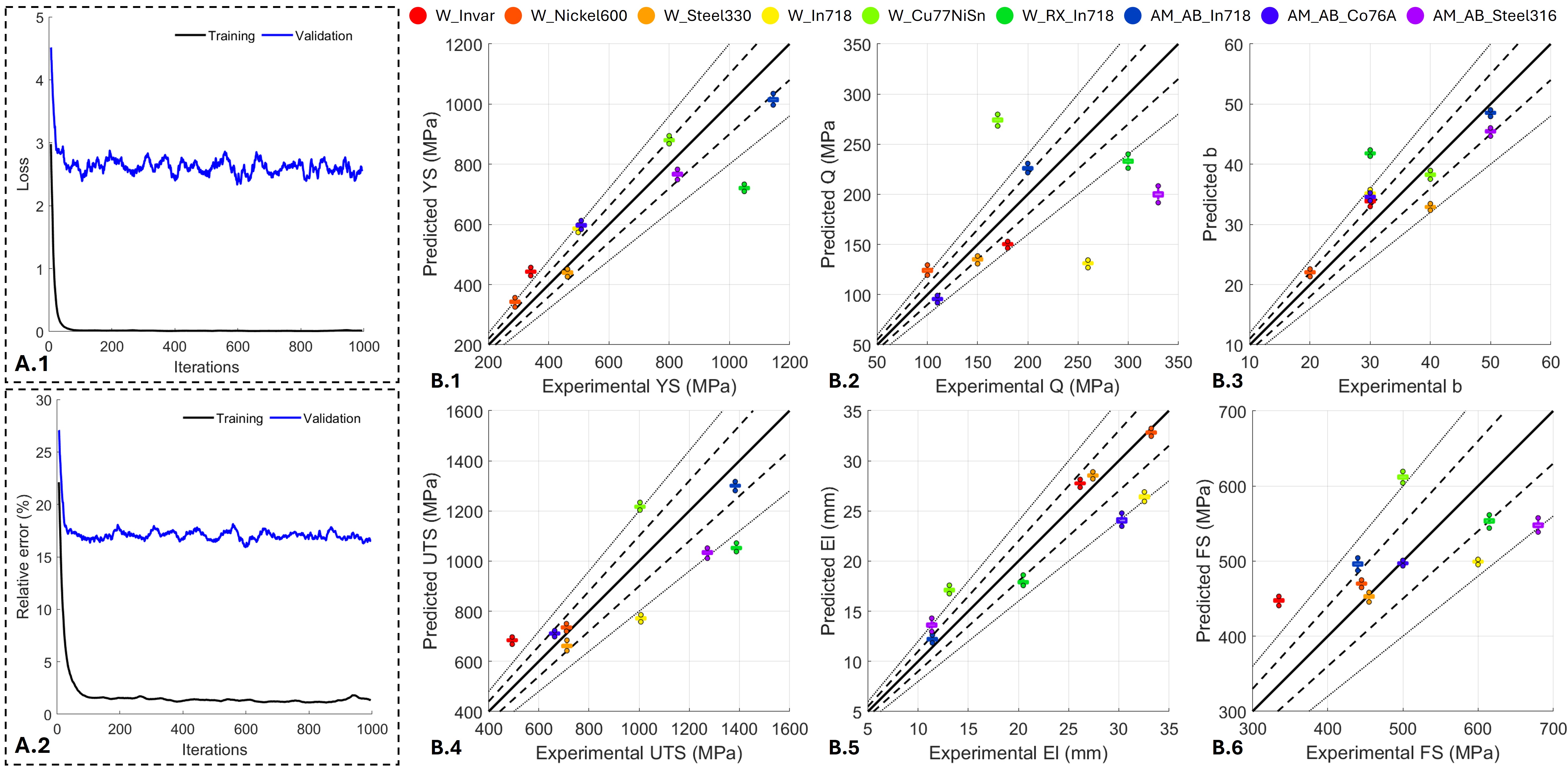}
    \caption{\textbf{Prediction of mechanical properties from plasticity.} \textbf{(A)} Training and validation metrics: \textbf{(A.1)} Loss and \textbf{(A.2)} relative error. \textbf{(B)} diagrams showing the predicted mechanical properties compared to the experimental properties: \textbf{(B.1)} yield strength, \textbf{(B.2)} Q hardening rate, \textbf{(B.3)} b hardening exponent, \textbf{(B.4)} ultimate tensile strength, \textbf{(B.5)} elongation at failure, \textbf{(B.6)} fatigue strength. The solid black line indicates a perfect match between experimental and predicted properties whereas the dashed and dotted lines delimit 10\% and 20\% of relative errors, respectively.}
    \label{fig:fit}
\end{figure}

\justify Following training, the mechanical properties were estimated from all supplementary samples from a simple deformation state. The different predicted properties are presented in Fig. \ref{fig:fit}(B.1-B.6) corresponding to the yield strength, hardening rate and exponent, ultimate tensile strength, elongation and fatigue strength, respectively. The solid black line in Fig. \ref{fig:fit}(B.1-B.6) shows identical values between the experimental and predicted properties. To illustrate the accuracies over the different properties and materials, dashed and dotted lines in Fig \ref{fig:fit}(B.1-B.6) delimit 10\% and 20\% of relative errors.  As the mechanical properties are estimated from $9 \times 9$ windows (153 $\mu$m $\times$ 153 $\mu$m) regions, multiple values are obtained for a given sample. The underlying distributions are summarized by boxes extending from 25\% to 75\% of the distribution and whiskers extending to 5\% and 95\%. Maximum and minimum values are indicated with circular markers. The prediction relative errors are also provided in Table \ref{tab:acc} in the supplementary materials and are reported in terms of lowest and highest relative errors. Most of the predicted properties fall within the 20\% relative error range with the exception of a few points. We discuss the method below, including the associated uncertainties and how it can be improved to enhance prediction accuracy and further extended it to other mechanical property predictions.

\justify The approach detailed in this study provides an accelerated framework for characterizing mechanical properties. Although the time gain is fairly limited when it comes to monotonic properties, significant gain in time is achieved for fatigue properties once the model is trained. To obtain the fatigue strength following the approach detailed in the methods section, on average, the samples were mechanically tested for 20 days. With the proposed method, the fatigue properties can be estimated within 17 hours, thereby achieving a 30-fold acceleration compared to the conventional testing method. The times relative to the different steps are listed in Table \ref{tab:time}. Using more advanced SEM technologies, the imaging time for the region size considered in this study ($960\times 630\:\mathrm{\mu m}$) could be completed within a few tens of minutes \cite{Black2023}. In addition, HR-DIC computation time can be optimized with dedicated GPU clusters to obtain a potential 100-fold acceleration to determine fatigue properties. Finally, since such measurements are performed locally, testing functionally graded specimens could be implemented, enabling the evaluation of multiple compositions and microstructures within a single specimen \cite{bean2025acceleratedfatiguestrengthprediction}.

\begin{table}[]
    \centering
    \caption{Estimated time for the different steps of the conventional and the proposed method.}
    \label{tab:time}
    \begin{tabular}{|c|c|c|}
        \hline
         & Proposed method & Fatigue testing \\
         & (Accelerated) & (Conventional) \\
         \hline
         Specimen preparation & $\sim$ 1.5 hours & $\sim$ 2 hours \\ 
         Mechanical testing & $\sim$ 1 hour & $\sim$ 20 days \\ 
         SEM Imaging & $\sim$ 4.5 hours & / \\ 
         HR-DIC computation & $\sim$ 12 hours & / \\ 
         \hline
    \end{tabular}
\end{table}

\justify The proposed approach involves measuring, encoding, and mapping the latent space features of local plasticity that develop during elementary loading. Three material-based hypotheses are considered: (i) a relationship exists between surface plasticity (and its heterogeneity) and macroscopic properties; (ii) elementary loading (and the measure of plasticity) can provide insights into macroscopic properties; and (iii) the encoding and mapping of latent space can accurately describe plasticity. The first two points are discussed in the next section, while the last point is addressed in the last section of the discussion.

\subsection*{Relationships between plasticity during elementary loading and macroscopic properties}

\justify The mechanisms governing monotonic macroscopic properties involve complex plastic deformation processes, including dislocation glide and dislocation interactions that can differ from those observed during early monotonic plastic deformation. In the example of fatigue properties, the controlling mechanisms at the small scale relate to slip irreversibility \cite{Mughrabi2009}, which leads to crack initiation, and crack propagation rate, ultimately resulting in specimen fracture. Interestingly, this fundamental mechanism, i.e. slip irreversibility, is directly influenced by the localization of plasticity within the material \cite{Stinville2022, Mughrabi2009, STINVILLE2020172}. A direct correlation exists between slip irreversibility and the intensity of plastic deformation events that develop early during monotonic loading \cite{annurev:/content/journals/10.1146/annurev-matsci-080921-102621,Stinville2022}. Materials that exhibit highly localized plasticity during early monotonic deformation have low fatigue ratio (fatigue strength divided by yield strength), and vice versa. It was demonstrated that the statistic analysis of the intensity of plastic deformation events by itself can be used to inform fatigue strength \cite{Stinville2022}. Here, predictions are generalized for a large set of macroscopic properties by considering additional information within the HR-DIC measurements. 

\justify Monotonic macroscopic properties are also embedded within the characteristics of metal plasticity. For example, the macroscopic yield strength of a material has been observed to correlate with both the intensity and length of slip events, the material's propensity for cross-slip and the events connectivity (slip transmission). Macroscopic yield strength is related, for instance, to the number of dislocations within the slip plane, which directly influences the slip intensity, and by the dislocation pile-up at interfaces such as grain boundaries \cite{HULL2011171, Andani2020, ANDANI2022117613, doi:https://doi.org/10.1002/9781119296126.ch173, doi:10.1126/sciadv.abo5735}. Dislocation dynamics simulations \cite{CHAKRAVARTHY2010625, CLEVERINGA1999837} show that yield strength scales with predictions from analytical continuum pile-up models and the number of dislocations involved in each slip event \cite{refId0, CHAKRAVARTHY2010625}. All of these characteristics are statistically captured from full-field HR-DIC measurements \cite{annurev:/content/journals/10.1146/annurev-matsci-080921-102621}.

\justify Furthermore, the macroscopic hardening rate of a material directly influences the development of plasticity during loading. The number and intensity of slip events, their connectivity through slip transmission, the number of activated slip systems, and the evolution of these features throughout deformation, captured by HR-DIC, are all directly correlated with the strain rate. Additionally, HR-DIC measurements provide more comprehensive information beyond slip intensity by capturing strain components and local rotations \cite{STINVILLE2020110600} (see Fig. \ref{fig:hrdic}). These measurements reflect lattice rotation and expansion associated with both elastic deformation and non-localized plasticity \cite{STINVILLE2020110600}, including isolated dislocations and geometrically necessary dislocations. Notably, lattice rotation and expansion have also been observed to correlate with the macroscopic hardening rate \cite{BITTENCOURT2018169}.

\justify Finally, ultimate tensile strength and elongation to failure are also closely linked to plasticity, as necking occurs when the material becomes "saturated" with plastic deformation or when significant strain accumulation takes place at grain boundaries. These properties are further influenced by the strain hardening rate. 

\justify Building on these previous considerations, it is not surprising to observe relatively good predictions (within 20\% error) from the developed model, even with a very small training set of alloys. Interestingly, the properties with the largest inaccuracies are related to the hardening rate, which depends not only on the distribution and characteristics of slip but also on non-localized plasticity (e.g., lattice rotation and expansion). In the current model, although the encoding of lattice rotation and strain maps has been developed, they have not been used to predict the mechanical properties at this stage. In addition, since our elementary loading and measurements capture plasticity only up to strain levels below a few percent, they do not fully capture how plasticity evolves (saturates) at higher deformation levels. This highlights the importance of selecting appropriate elementary loading conditions and all modality provided by the HR-DIC measurements. By including additional loading sequences, such as a few loading cycles or "creep" loading, as schematically illustrated in Fig. \ref{fig:principle}, extending the deformation to higher levels, using different loading strain rates, and considering all encoded HR-DIC modalities, a deeper insight into plasticity and its evolution should be achieved, and consequently the accuracy of macroscopic property predictions should be greatly improved.

%Nevertheless, advancements in mechanical loading automation and electron scanning technology can significantly reduce the time required for HR-DIC measurements. For instance, these measurements can now be completed within a few minutes for square millimeter fields of view analyses. A notable example is the development of multi-beam technology in scanning electron microscopy, which has accelerated HR-DIC measurements by two orders of magnitude \cite{Black2023}.

\subsection*{Heterogeneity of plasticity localization and mechanical properties prediction} % Metal behavioral genomics

\justify Different methods have then been proposed in the literature to extract plastic deformation events from HR-DIC data. These various methods rely on physics-based assumptions about the morphologies of deformation events \cite{Vermeij2023,Charpagne2020,Hu2023,Ni2024,Bean2025} and need to be adapted when considering new materials, microstructures or deformation mechanisms. Such approaches allow to focus on the characteristics of individual events (intensity, length, orientations) which are generally summarized by scalar values. Therefore, information is lost through these physics-based analyses by focusing on only a few physical characteristics of plasticity. As an alternative, variational autoencoders provide an efficient methodology to fully describe fields by creating a low-dimensional representation without any physics-based hypothesis \cite{Calvat2025,Biswas2023,Valleti2024}. This self-supervised method is suitable for various materials and deformation mechanisms without any manual annotation or physical consideration. Compared to the segmentation and extraction approaches, the VAE strategy is a lossless method and can theoretically extract all the information contained in the input fields \cite{Goodfellow2016}. 

%\justify The low-dimensional representations were subsequently used to train a CNN architecture to predict the mechanical properties listed in table \ref{tab:props}. Pragmatically, all the $9 \times 9$ windows were indifferently used to predict the set of mechanical properties. However, fatigue properties are mainly controlled by the most intense events\cite{Stinville2022}, the higher the plastic deformation intensity, the worse the fatigue properties. As the properties predictions is made of a limited region and therefore may not contain the most intense events, the prediction accuracy is expected to mostly depend on the capture of these high-intensity deformation events. Out of training, a multitude of mechanical properties are predicted from a single HR-DIC map and therefore requires adapted computation method to estimate the likelihood a predicted property.

\justify More importantly, the proposed method enables the characterization of plasticity heterogeneity through the mapping of latent space features that encode plasticity. This novel concept represents a shift from the current data-driven materials science approaches, which typically rely on discrete descriptors and often lose spatial information, i.e. material heterogeneities. For instance, conventional approaches such as graph networks or regression/correlation-based methods \cite{MANGAL2018122,MANGAL20191,Thomas2023,Dai2021} are commonly used to predict macroscopic or local properties but fail to preserve the full extent of material heterogeneity. Preserving information related to the heterogeneity is of critical importance in metal research, as metals often exhibit significant small-scale variability over large fields of view, which ultimately governs their macroscopic properties \cite{SU201940}. Here, we consider large regions (153 $\mu$m $\times$ 153 $\mu$m windows) to capture these heterogeneities. The size of the region is a critical hyperparameter, increasing the region size would consequently improve the representativeness and therefore enhance the accuracy of predictions. Building a fixed-size input CNN architecture defines a simple framework, but suffers from an imbalance of accuracy between different microstructures. While a given region may be statistically representative for a fine-grained microstructure, the same region size may not be representative for large-grained microstructures such as those produced by additive manufacturing.

\subsection*{Structure of the latent space}

\begin{figure}[h!]
    \centering
    \includegraphics[width=1\linewidth]{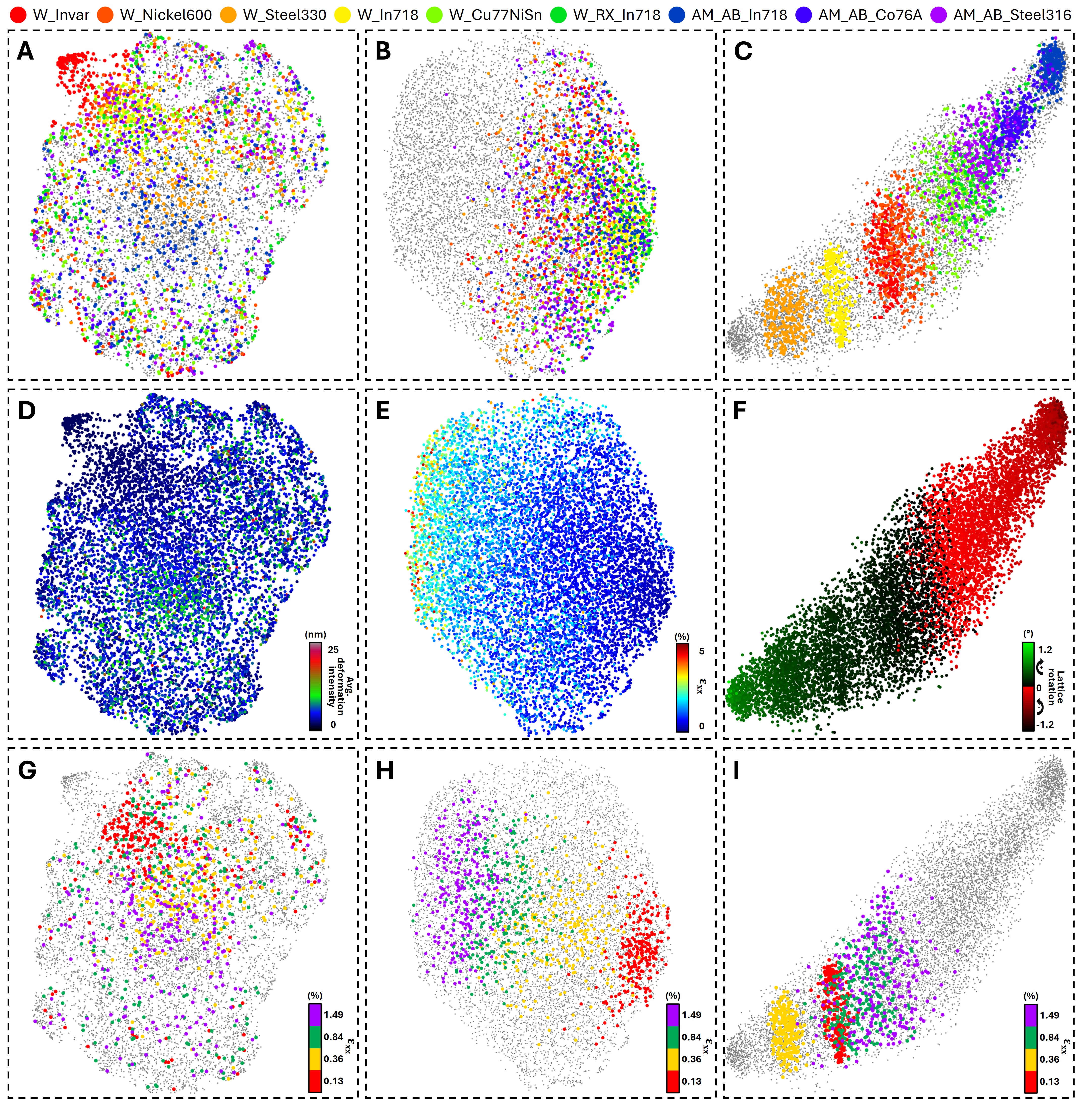}
    \caption{\textbf{Representation of materials and their plasticity within the latent space.} Uniform manifold approximation and projection applied to different latent spaces and illustrating the partitioning of materials: \textbf{(A)} plastic deformation intensity, \textbf{(B)} longitudinal strain and \textbf{(C)} lattice rotation. Average values evolution in the different latent spaces: \textbf{(D)} plastic deformation intensity, \textbf{(E)} longitudinal strain and \textbf{(F)} lattice rotation. Variations associated to the increase of macroscopic plastic strain of W\_In718 observed in the different latent spaces: \textbf{(G)} plastic deformation intensity, \textbf{(H)} longitudinal strain and \textbf{(I)} lattice rotation.}
    \label{fig:umap}
\end{figure}

\justify The $\mu$ low-dimensional representations have been produced by encoding the complete VAE dataset (training, validation and testing) and the different HR-DIC modalities. These low-dimensional representations have been transformed into 2D representations using uniform manifold approximation and projection (UMAP) \cite{Mcinnes2018}. The same parameters have been used for all the projections: minimum distance of 0.02, 150 neighbors and spread of 3. Firstly, the repartition of the different materials at early plasticity (lowest macroscopic plastic strain) in different latent spaces are shown in Fig. \ref{fig:umap}(A,B,C) and corresponding to the plastic deformation intensity, longitudinal strain ($\varepsilon_{xx}$) and lattice rotation, respectively. The low-dimensional representations at higher macroscopic strains have been considered for UMAP but are first represented as gray markers. While the distribution of the early plasticity representations over the manifold is homogeneous for the plastic deformation intensity and the lattice rotations, when it comes to longitudinal strain, only half of the manifold seems to correspond to early plasticity. For lattice rotations, low-dimensional representations are grouped into clusters and might be related to rigid body rotations which were not corrected in this study. In addition, Fig. \ref{fig:umap}(D,E,F) display the average values that are associated with the low-dimensional representations of all levels of macroscopic plastic strains. The projected manifold shows well identified regions of high and low average values. Interestingly, the plastic deformation intensity projection given in Fig. \ref{fig:umap}(D) presents several disconnected low-intensity regions. As an example, Fig. \ref{fig:umap}(G,H,I) illustrate how the W\_In718 low-dimensional representations vary over the manifolds with increasing plastic strain. Regardless of the manifold considered, the representations are initially grouped in a given region at early plasticity. These representations move in the latent space with increasing plasticity and tend to spread out as the heterogeneities between the reduced regions increase.

\justify The plots in Fig. \ref{fig:umap} illustrate the underlying structure of the latent space. Interestingly, clusters or gradients are observed based on both the physical descriptors of plasticity, such as $\varepsilon_{xx}$ strain and lattice rotation, as well as the material types. This highlights that the structure of the latent feature space and the associated processing effectively capture key physical information related to plasticity, as described by the HR-DIC modalities and their fields.

%Ultimately, such representations can be extended to other distribution characteristics (variance, skewness, etc) but also spatial descriptors (length of events, event spacing, etc). The property prediction framework can be extended to evaluate the impact of a given window on the mechanical properties. For instance, an intense localization of plasticity leads to poor fatigue properties\cite{Stinville2022}, an appropriate region of high-intensity events can be defined in the latent space and shall be avoided to maximize fatigue properties. This consideration only constrains the localization of plasticity. From a microstructure design perspective, additional developments are consequently needed to predict quantitatively the localization of plasticity from a given microstructure.

\subsection*{Extending the database} \label{sec:extending}
% Performances on unseen mechanisms

\justify Extending the database by adding more microstructures, chemical compositions and loading conditions (i.e. elementary loadings) would greatly improve the capability of the VAE to encode plasticity. However, as the database is growing, it becomes very time consuming (especially with conventional GPU workstations) to train the VAE and the CNN architectures from scratch each time. This is one of the important limitations on the proposed approach. 

\begin{figure}[htbp]
    \centering
    \includegraphics[width=1\textwidth]{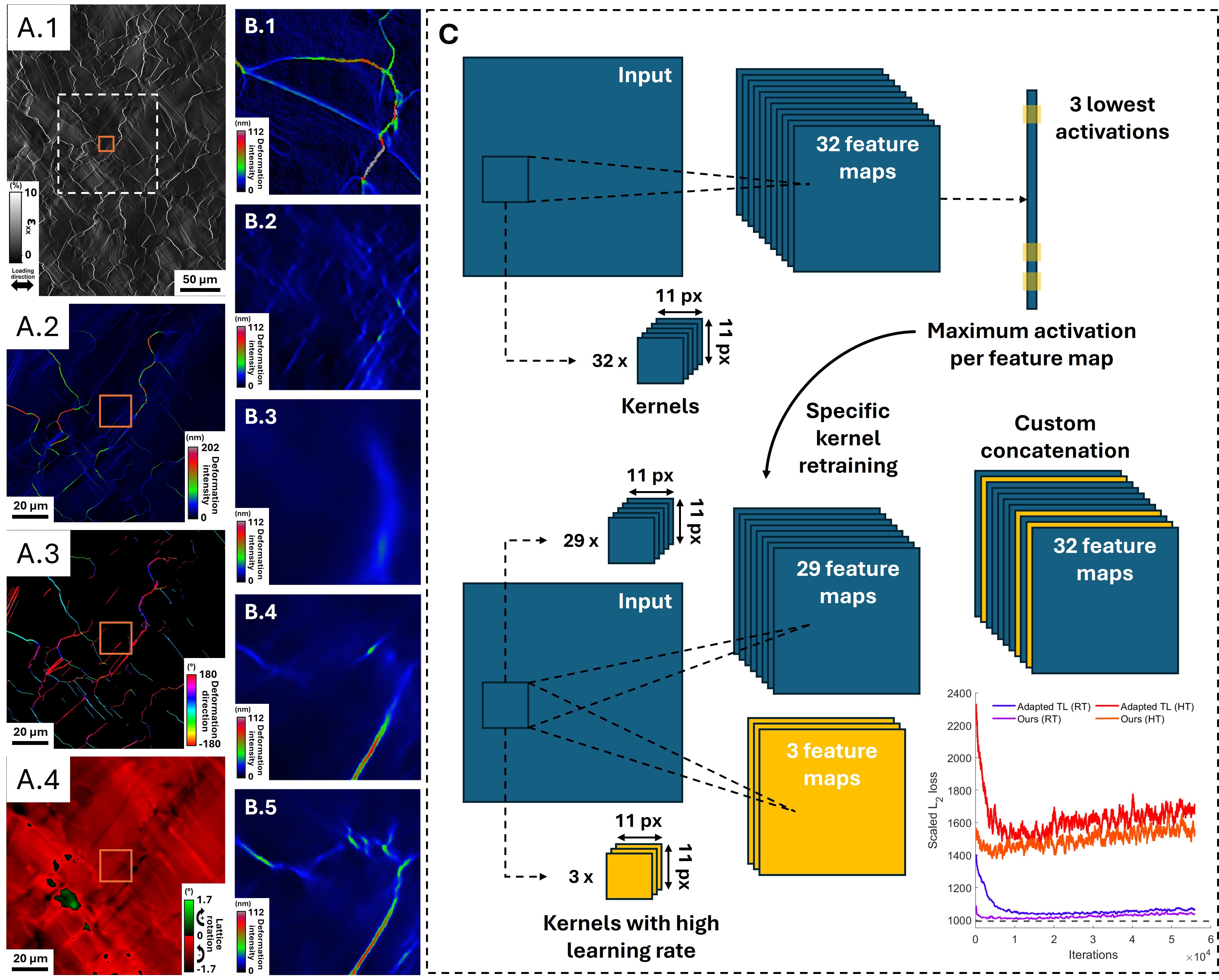}
    \caption{\textbf{Dataset augmentation with new experimental data.} Invar deformed at 500$^\circ$C at 0.90\% of macroscopic plastic strain and associated \textbf{(A.1)} longitudinal strain, \textbf{(A.2)} plastic deformation intensity, \textbf{(A.3)} plastic deformation direction and \textbf{(A.4)} lattice rotation. As an example, a 256 $\times$ 256 window with an orange border is shown on the different HR-DIC modalities. \textbf{(E)} Proposed method to partially retrain the VAE architecture. \textbf{(B.1)} Original plastic deformation intensity and encoded/decoded through VAE architectures trained on \textbf{(B.2)} wrought materials, without augmentations and all materials deformed at RT: \textbf{(B.3)} without augmentations, \textbf{(B.4)} with reflections and \textbf{(B.5)} with reflections and masking. \textbf{(C)} Our custom retraining method based on the activation when processing the dataset and scaled $L_2$ loss comparison.}
    \label{fig:invar_500}
\end{figure}

\justify As an example, the HR-DIC modalities from the deformation of W\_Invar at 500$^\circ$C are considered. The associated longitudinal strain ($\varepsilon_{xx}$), plastic deformation intensity, direction and lattice rotation are shown for a reduced region of interest in Fig. \ref{fig:invar_500}(A.1,A.2,A.3,A.4), respectively. Compared to the different HR-DIC modalities presented in Fig. \ref{fig:hrdic}, at this temperature, Invar shows grain boundary sliding \cite{Bean2025} which is a different deformation process in comparison to slip that drastically induces different plastic fields that are not yet part of the database. Fig. \ref{fig:invar_500}(B.2,B.3,B.4,B.5) show the reconstructions of a plastic deformation intensity window using the VAE with various augmentations, the original version of which is given in Fig. \ref{fig:invar_500}(B.1). All reconstructions are associated with VAE trained on HR-DIC data obtained only at room temperature (deformation by slip). The reconstructions shown in Fig. \ref{fig:invar_500}(B.2,B.4,B.5) correspond to the trainings presented in the results section: without augmentation, with reflections, and with all considered augmentations, respectively. An additional training was carried out using only wrought materials (W\_Invar, W\_Nickel600, W\_Steel330, W\_In718, W\_Cu77NiSn and W\_RX\_In718) without any augmentation, the reconstruction of which is shown in Fig. \ref{fig:invar_500}(B.2). Using only wrought materials limits the database to straight events and reduces the database size.  As a consequence the plastic deformation intensity is poorly reconstructed. With the full RT database without augmentation, the latent space focuses on describing the most intense grain boundary sliding event. When augmentations are considered, the reconstruction improves even if grain boundary sliding is not yet part of the database, with the best reconstruction being when all augmentations are considered (Fig. \ref{fig:invar_500}(B.5)). However, the morphology of grain boundary sliding gets oversimplified upon reconstruction. 

\justify First, the general idea associated to transfer learning \cite{Weiss2016} (TL) has been applied to the VAE approach to include grain boundary sliding, i.e. the W\_Invar deformed at high-temperature, to the database without retraining the VAE architecture from scratch. For a more comprehensive visualization of the data associated with the W\_Invar deformed at high-temperature, the longitudinal strain component, $\varepsilon_{xx}$, are displayed in Multimedia Component 1. Under these circumstances, the first convolutional layers are assumed to be able to process the features associated with grain boundary sliding. Only the last layers of the \textit{Encoder} are reinitialized and trained again with the complete dataset (all materials previously considered and the data associated with the W\_Invar deformed at high temperature). As an alternative, we propose a new retraining method to incorporate new deformation mechanisms to the database which is schematically illustrated in Fig. \ref{fig:invar_500}(C). First, the full dataset is processed through the complete VAE architecture (both \textit{Encoder} and \textit{Decoder} networks) to identify the highest activations for all depths of the CNNs (only shown for the first set of convolution kernels). These results help to identify the relevance of each convolution kernel and, by opposition, to identify the 10\% less relevant (lowest activations) to be considered for retraining (3 out of 32 kernels to be retrained for the first convolution layer of the \textit{Encoder}). To improve training stability, the layers to be retrained are not directly removed but their output is scaled by a factor $\left(1 - \gamma\right)$ and the new layers by a factor $\gamma$ which is slowly increased to 1. The losses associated to the adapted TL and our custom retraining method are given in Fig. \ref{fig:invar_500}(C). Both of these methods lead to an RT validation loss higher than the optimal validation loss previously obtained in Fig. \ref{fig:loss}(A) and materialized by the horizontal black dashed line. Using our training method allows to achieve lower validation losses for both slip and grain boundary sliding-dominated plasticity compared to the adapted TL method. The optimal training point is also obtained with half the number of iterations (considering the HT validation loss). 

\section{Conclusion}

The proposed approach enables rapid prediction of mechanical properties, accelerating material design and offering significant time savings compared to conventional testing. Additionally, the method relies on local full-field measurements, enabling it to capture the variability in mechanical properties caused by material heterogeneity. It also can facilitate a fundamental understanding of the effects of microstructure on mechanical properties. This approach has the potential to be extended to other mechanical properties, such as creep and fatigue-creep, or to loading under extreme conditions (e.g., high temperature or high strain rate) by incorporating additional elementary loading modes (see Fig. \ref{fig:principle}). Incorporating these additional loading conditions and accounting for all HR-DIC modalities (which is not addressed here) will further enhance predictive accuracy. Moreover, although not detailed here, the present approach can offer a fundamental understanding of the characteristics of plasticity at a small scale on mechanical properties. As predictive accuracy improves with more data, interpretability tools such as permutation importance, occlusion sensitivity, saliency maps, and others will help identify the key characteristics of plasticity and heterogeneity that govern each macroscopic property. Finally, by predicting mechanical properties from maps of encoded plasticity, the variability in mechanical properties can be evaluated from one region to another, thereby establishing the scatter of properties coming due to material heterogeneity. For example, maps of the encoded plasticity values are shown in Fig. \ref{fig:Intro}(B) and show variation from one region to another. Similar maps are generated for the macroscopic properties, providing insight into the variability coming from microstructural heterogeneity. This offers a unique opportunity to identify beneficial microstructural features for the design of new microstructures and alloys.

%At this point, the plastic deformation events do not belong to different classes during encoding, but their intensity may vary at different rates, as noted for the 'zigzag' patterns \cite{Bean2025}. One solution might be to separate the encoding of the deformation mechanisms into different latent spaces for further use to enhance the accuracy of mechanical properties prediction. Furthermore, the VAE and CNN architectures were trained separately, but they could -- in principle -- be trained simultaneously to optimize the reconstruction for characteristics that control the mechanical properties.

\section{Experimental Section}

\subsection*{Materials and sample preparation}

\justify A total of 9 FCC materials were investigated in this study, including wrought and additively manufactured materials, and are listed in Table \ref{tab:materials}. \textbf{W}, \textbf{AM}, \textbf{RX}, \textbf{AB} stand for Wrought, Additively Manufactured, Recrystallized and As-Built, respectively. W\_RX\_In718 was annealed at 1050$^{\circ}$C for 30 minutes to generate a nearly random texture, followed by 8 hours at 720$^{\circ}$C, forming $\gamma$' precipitates. AM\_AB\_In718, AM\_AB\_Co76A and AM\_AB\_Steel316 were produced with a Formalloy L2 Directed Energy Deposition (DED) unit utilizing a 650 W Nuburu 450 nm blue laser capable of achieving a 400~$\mu$m laser spot size; Argon was used as the shielding and carrier gas.

\begin{table}[h!]
    \centering
    \setlength{\tabcolsep}{3pt}
    \caption{Chemical composition, in wt.\%, of the alloys considered in this study. Values denoted by a \textsuperscript{m} correspond to maximum values related to specifications.}
    \label{tab:materials}
    \begin{small}
        \begin{tabular}{|c||c|c|c|c|c|c|c|c|c|c|c|c|c|c|c|}
            \hline
            \textbf{Alloy} & Fe & Ni & Cu & Co & Cr & Nb & Mo & Mn & Ti & Al & Sn & Si & C & Other \\
             \textbf{Denomination} & ($\%$) & ($\%$) & ($\%$) & ($\%$) & ($\%$) & ($\%$) & ($\%$) & ($\%$) & ($\%$) & ($\%$) & ($\%$) & ($\%$) & ($\%$) & ($\%$) \\
             \hline
             W\_Invar & bal. & 38 & / & 0.5 & 0.25 & / & / & 0.6 & / & 0.1 & / & / & 0.05 & 0.1 \\
             \hline
             W\_Steel330 & bal. & 34--37 & 1\textsuperscript{m} & / & 18--20 & / & / & 2\textsuperscript{m} & / & / & / & 1.5\textsuperscript{m} & 0.08\textsuperscript{m} & 0.3 \\
             \hline
             W\_Nickel600 & 6--10 & bal. & / & / & 14--17 & / & / & 1\textsuperscript{m} & / & / & / & 0.5\textsuperscript{m} & / & / \\
             \hline
             W\_In718 & 23.5 & bal. & / & / & 17--21 & 4.75--5.5 & 2.8--3.3 & 0.35\textsuperscript{m} & 0.65--1.15 & 0.2--0.8 & / &  0.35\textsuperscript{m} & 0.08\textsuperscript{m} & / \\
             \hline
             W\_Cu77NiSn & 0.5\textsuperscript{m} & 14.5--15.5 & bal. & / & / & 0.1\textsuperscript{m} & / & / & / & / & 7.5--8.5 & / & / & 0.7 \\
             \hline
             W\_RX\_In718 & 17.31 & bal. & / & 0.14 & 17.97 & 5.4 & / & / & 1 & 0.56 &  / & / & 0.023 & / \\
             \hline
             AM\_AB\_In718 & 18.77 & bal. & 0.02 & 0.07 & 18.88 & 5.08 & 3.04 & 0.04 & 0.96 & / & / & 0.08 & 0.036 & / \\
             \hline
             AM\_AB\_Co76A & / & / & / & bal. & 29 & / & 6 & / & / & / & / & / & 0.12 & 0.1 \\
             \hline
             AM\_RX\_Steel316 & bal. & 12 & / & / & 18 & / & 2 & / & / & / & /  & / & 0.03 & 1 \\
             \hline
        \end{tabular}
    \end{small}
\end{table}

\begin{figure}
    \centering
    \includegraphics[width=.7\linewidth]{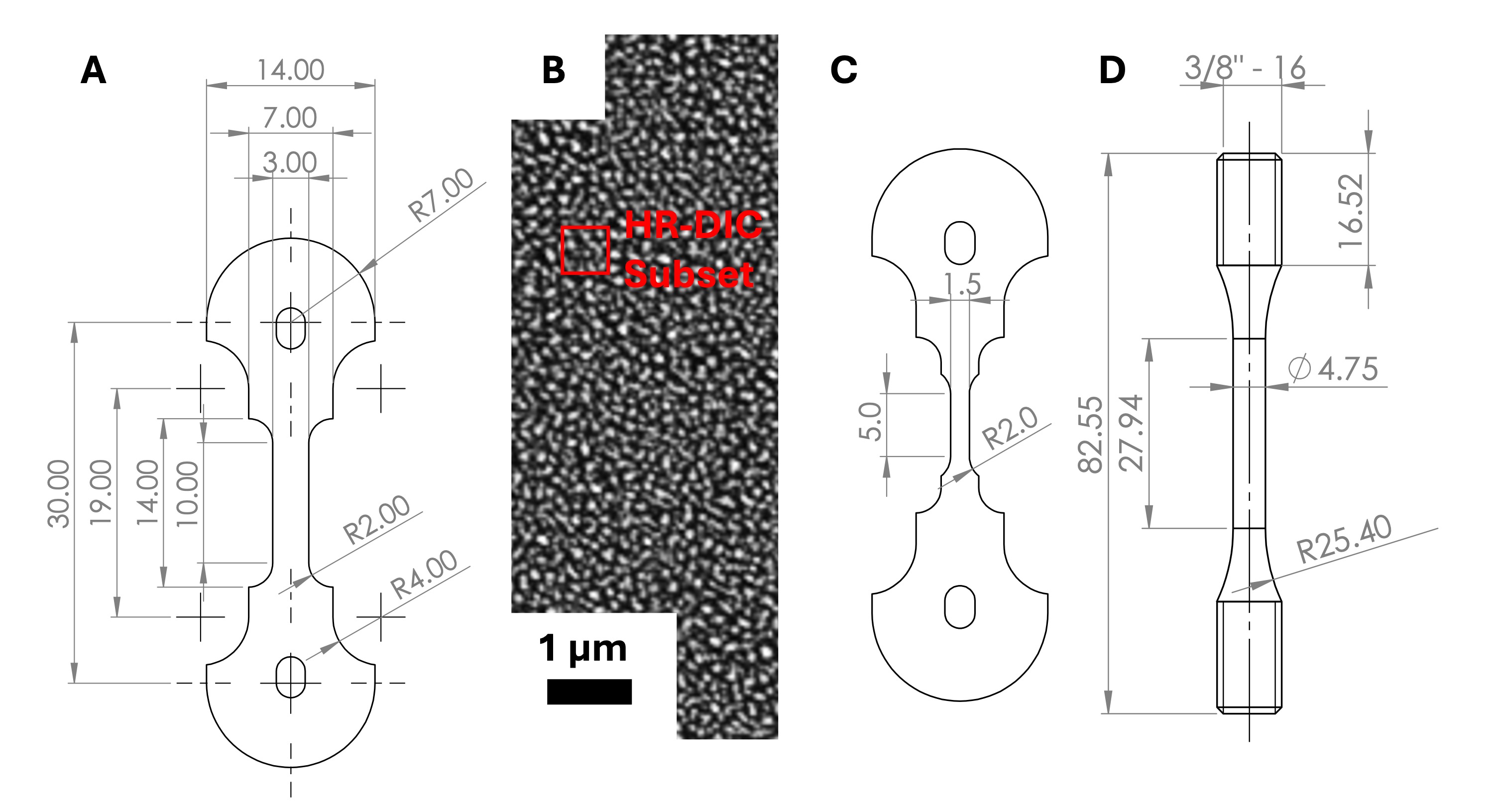}
    \caption{\textbf{(A)} HR-DIC sample geometry and \textbf{(B)} associated nanometer scale speckle. The red box displays the subset size used for the HR-DIC calculation. \textbf{(C)} Modified sample geometry for optical DIC measurements to obtain macroscopic monotonic mechanical properties. \textbf{(D)} Specimen geometry for mechanical fatigue testing. Dimensions are provided in millimeters, with the exception of the thread in (D) which is given in inches.}
    \label{fig:sample}
\end{figure}

\justify HR-DIC samples were machined by EDM (electrical discharge machining) as flat dogbone samples of gauge section $1 \times 3\:\mathrm{mm^2}$ and the corresponding geometry is presented in Fig. \ref{fig:sample}(A). Macroscopic fatigue specimens were machined by CNC following the geometry provided in Fig. \ref{fig:sample}(D). All samples were mechanically polished using abrasive papers up to 1200 grit followed by diamond suspension down to $3\:\mathrm{\mu m}$. Additionally, the HR-DIC samples were finished using a $50\:\mathrm{nm}$ colloidal silica suspension either manually for 20 minutes or on a vibratory polisher for 24 hours. Layers of $3\:\mathrm{nm}$ of Titanium followed by $10\:\mathrm{nm}$ of Silver were deposited on the samples using an AJA Sputter Coater. The silver top layer was reconfigured into particles by immersion in a 1\% saltwater solution for 2 hours following procedures detailed in references\cite{Kammers_2013a, Montgomery2019}. This procedure was applied for all materials except the W\_Cu77NiSn -- due to corrosion issues -- which has been heat treated at $275^{\circ}\mathrm{C}$ for 90 minutes under vacuum ($10^{-4}\:\mathrm{Torr}$).

\subsection*{Mechanical Testing and Microscopy}

\justify Fatigue testing was carried out in an incremental manner up to 500000 cycles and the first stress was chosen to be 50\% the yield strength. Fatigue testing was performed at a stress ratio $R_{\sigma}$ of 0.1 and at $2\:\mathrm{Hz}$ onto an Instron 8500 and an Instron 8800 with $100\:\mathrm{kN}$ force cells. If no fracture was observed, the stress was raised by 10\% of the yield strength and this operation was reiterated until fracture. Following the adopted incremental testing, Fatigue Strength (FS) was identified as the highest stress of run-out before failure. The identified mechanical properties are given in Table \ref{tab:props}.

\begin{figure}
    \centering
    \includegraphics[width=1\linewidth]{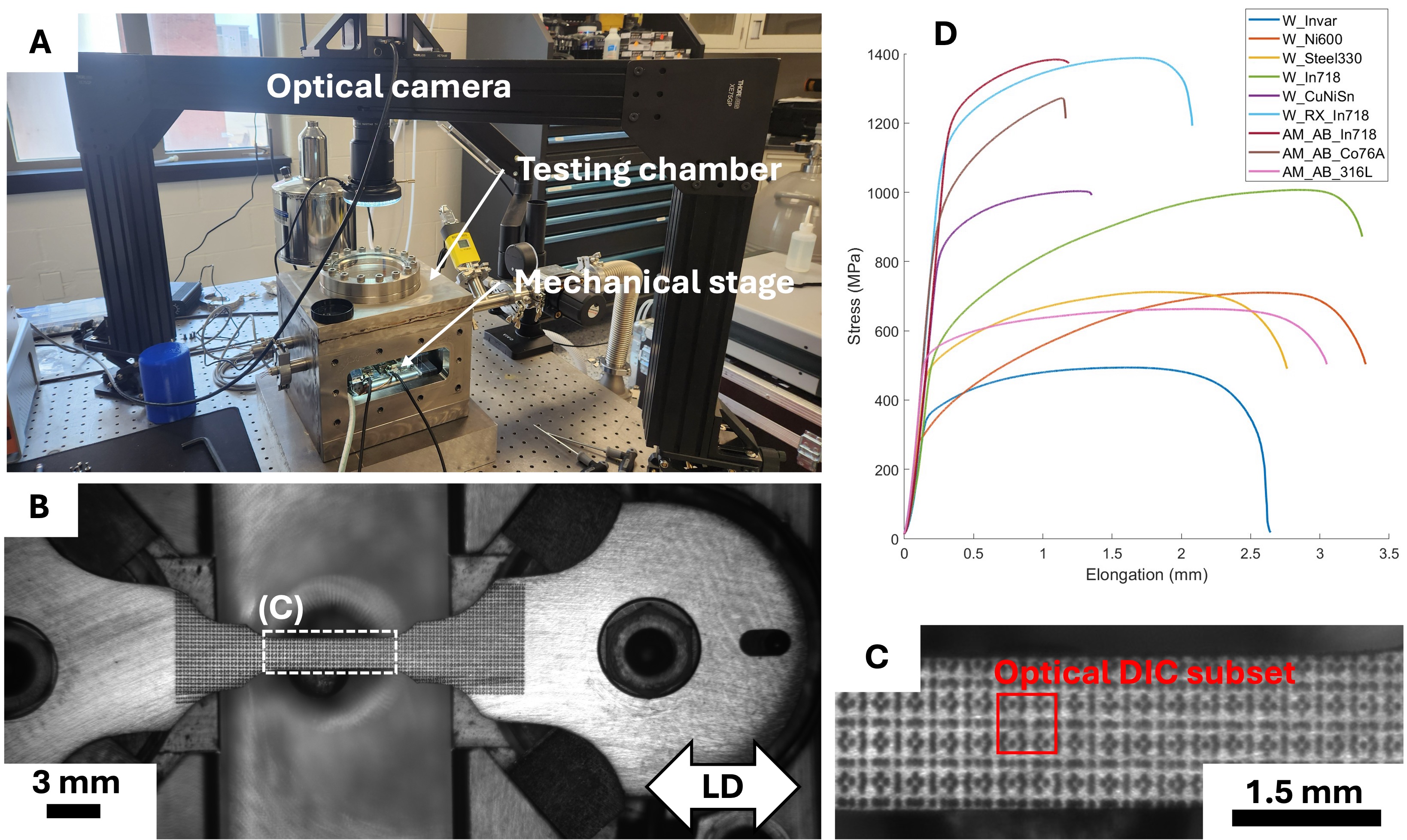}
    \caption{\textbf{(A)} Experimental setup to extract macroscopic monotonic properties. Conventional tensile tests were performed with optical DIC measurement. A high-resolution optical camera is used to capture images of the specimen surface during deformation. \textbf{(B)} Example of a high-resolution image used for optical DIC computations and \textbf{(C)} speckles produced by laser engraving. \textbf{(D)} Monotonic behaviors of the different materials considered in this study.}
    \label{fig:testing}
\end{figure}

\justify On the other hand, HR-DIC and optical DIC mechanical testing was performed using a $5\:\mathrm{kN}$ NewTec MT1000 at a rate of $2\:\mathrm{N\cdot s^{-1}}$ corresponding to quasi-static loading conditions. All HR-DIC samples were interrupted at increasing levels of macroscopic plastic strain to collect images. High temperature tests were performed under a vacuum chamber to minimize surface oxidation. Prior to optical DIC, speckles were produced by laser engraving using an OMTech 30W laser with a wavelength of $1064\:\mathrm{nm}$. Optical DIC tests were then performed up to fracture to identify the relevant mechanical properties and images were captured every 5 seconds using a Pixelink PL-D7912MU-T camera (Fig. \ref{fig:testing}). Yield Strength (YS), Ultimate Tensile Strength (UTS), Elongation were identified from the tensile curves. These mechanical properties are listed in Table \ref{tab:props}.

\justify The high-resolution imaging was performed on a Thermofischer Scios 2 Dual Beam SEM/FIB for high-resolution digital image correlation (HR-DIC). SEM images were taken after unloading at the different macroscopic plastic strain states. To mitigate distortions errors related to SEM imaging, SEM parameters were selected in accordance with guidelines from Kammers and Daly \cite{Kammers_2013b}, Stinville \textit{et al.}\cite{Stinville_2015a}, and Mello \textit{et al.}\cite{Mello2017}. The imaged area, $960\times 630\:\mathrm{\mu m}$, consists in a grid of 7 by 7 secondary electron images with $15\%$ overlap between them. The 6144 by 4096 pixels images were acquired at an accelerating voltage of $10\:\mathrm{kV}$, current of $0.8\:\mathrm{nA}$, a dwell time of $10\:\mathrm{\mu s}$ per pixel and at a working distance of $5\:\mathrm{mm}$. 

\justify Following optical DIC, longitudinal strain was extracted by averaging the strain within the gauge length of the samples. Young's modulus was then identified from the strain stress curve and used within numerical modeling. All considered materials were attributed a Poisson ratio $\nu$ of 0.3. For these tensile behaviors, isotropic nonlinear hardening was modeled using Eq. \ref{eq:hardening} according to monotonic loading in the elastoplastic regime\cite{Besson2009}. The hardening parameters $Q$ and $b$ were then manually identified and are listed in Table \ref{tab:props}. All strains and stresses listed in Table \ref{tab:props} are engineering values.

\begin{equation}
    \sigma = \sigma_y + Q(1 - e^{-b\varepsilon^p})
    \label{eq:hardening}
\end{equation}

\begin{table}[h!]
    \centering
    \setlength{\tabcolsep}{4pt}
    \caption{Mechanical properties of the alloys considered in this study.}
    \label{tab:props}
    \begin{tabular}{|c||c||c||c|c|c|c|c|c|c|c|c|c|}
        \hline
        \textbf{Alloy} & Training & Validation & \multicolumn{6}{c|}{Conventional properties}  \\
        \textbf{Denomination} & \textbf{$\varepsilon_{xx}^p$} & \textbf{$\varepsilon_{xx}^p$} & \textbf{YS} & \textbf{Q} & \textbf{b} & \textbf{UTS} & \textbf{El} & \textbf{FS} \\
        & (\%) & (\%) & (MPa) & (MPa) & & (MPa) & (mm) & (MPa) \\
        \hline
        W\_Invar & 0.17 / 0.28 / 1.00 / 1.98 & 0.34 & 340 & 180 & 30 & 494 & 26.15 & 335 \\
        \hline
        W\_Nickel600 & 0.34 / 0.51 / 1.16 / 1.63 & 0.34 & 287.6 & 100 & 20 & 710.3 & 33.24 & 445 \\
        \hline
        W\_Steel330 & 0.14 / 0.35 / 0.56 / 0.86 & 0.22 & 462.1 & 150 & 40 & 712.1 & 27.42 & 455 \\
        \hline
        W\_In718 & 0.13 / 0.36 / 0.84 / 1.49 & 0.34 & 497.4 & 260 & 30 & 1007.1 & 32.56 & 600 \\
        \hline
        W\_Cu77NiSn & 0.28 / 0.53 / 1.02 / 1.74 & 0.27 & 800 & 170 & 40 & 1003.5 & 13.12 & 500 \\
        \hline
        W\_RX\_In718 & 0.31 / 0.65 / 0.97 / 1.61 & 0.24 & 1049.7 & 300 & 30 & 1388.2 & 20.48 & 615 \\
        \hline
        AM\_AB\_In718 & 0.27 / 0.74 / 1.50 / 2.43 & 0.25 & 1145.9 & 200 & 50 & 1383.6 & 11.42 & 440 \\
        \hline
        AM\_AB\_Co76A & 0.26 / 0.61 / 1.12 / 1.82 & 0.28 & 827.6 & 330 & 50 & 1272.4 & 11.36 & 680 \\
        \hline
        AM\_AB\_Steel316 & 0.35 / 0.63 / 1.09 / 1.76 & 0.27 & 507.8 & 110 & 30 & 663.2 & 30.31 & 300 \\
        \hline
    \end{tabular}
\end{table}

\subsection*{High-Resolution Digital Image Correlation}

\justify The Heaviside-DIC technique, detailed in reference\cite{Valle2017}, was used to estimate the in-plane slip amplitude. It was performed using a proprietary software XCorrel and was computed with respect to the initial state. All the HR-DIC modalities were computed using a subset size of 35 $\times$ 35 pixels used with a step size of 3 pixels. HR-DIC computations were performed on a workstation equipped with NVIDIA Geforce RTX 4090 and NVIDIA Geforce RTX 3090. Additionally, the optical DIC was performed using the same software. Following computations, all maps were manually cleaned using ImageJ. 

\justify Different maps were obtained from HR-DIC measurements, including strain components $\varepsilon_{xx}$, $\varepsilon_{yy}$, $\varepsilon_{xy}$ and $\varepsilon_{yx}$, in-plane slip intensity \cite{STINVILLE2020110600}, and in-plane lattice rotation \cite{STINVILLE2020110600}. The strain components $\varepsilon_{xx}$, $\varepsilon_{yy}$, $\varepsilon_{xy}$ and $\varepsilon_{yx}$ represent the strain along the loading direction (horizontal), the transverse direction, and the shear components respectively. During deformation by slip, dislocations emerge at the specimen surface along crystallographic planes, forming slip traces \cite{Mughrabi2009}. Localized in-plane displacements occur across these slip traces \cite{BOURDIN2018307}, representing the relative displacement of the material on either side of the trace. The Heaviside-DIC method provides the amplitude of these slip displacements, as illustrated in Fig. \ref{fig:hrdic}(B). Additionally, using internal DIC parameters within the Heaviside-DIC framework \cite{STINVILLE2020110600}, the rotational field at each point can be obtained, excluding rotations induced by slip. The resulting maps correspond to the in-plane lattice rotation. Further details can be found elsewhere \cite{STINVILLE2020110600}.

\subsection*{HR-DIC data preparation}

\justify Prior to VAE trainings, the HR-DIC data (examples given in Fig. \ref{fig:hrdic}) were prepared. First, longitudinal strain and lattice rotation were preprocessed with Gaussian blur of standard deviation of 3 and 10, respectively. To improve the training of the VAE architectures, the longitudinal strain was multiplied by 200 and the plastic deformation intensity was divided by 25. Both plastic deformation direction and lattice rotation are angular values, however, due to the high values associated to the plastic deformation direction, these maps have been converted to their cosine and sine components. Additionally, a threshold on the plastic deformation intensity has been considered as shown in Fig. \ref{fig:hrdic}(C.1, C.2, C.3, C.4) and corresponding to 0.3 pixels (6.74 nm) associated to the HR-DIC resolution. On the other hand, the lattice rotation has not been treated as an angular value and has only been attributed a multiplication factor of 100. As shown in Fig. \ref{fig:aug}, when masking augmentation was considered, the deformation events were replaced by Gaussian noise to replicate the background.

\subsection*{VAE and CNN trainings}

\justify Within both VAE and CNN architecture for properties prediction, leaky ReLU (scale factor of 0.2) was used as activation functions after each convolution layer (and transposed convolution) to avoid the 'dying ReLU' issue \cite{Lu2019}. All weights were initialized using the Glorot initializer \cite{Glorot2010}. The first trainings associated to the VAE architectures were carried out with a batch size of 128 for up to 3,000 epochs and a learning rate of $5\cdot10^{-5}$ kept constant during training (identical for both the \textit{Encoder} and the \textit{Decoder}). The HR-DIC dataset corresponds to 9,216 $256 \times 256$ windows and the dataset was split into training, validation, and testing portions of 70\% / 20\% / 10\%, respectively. All HR-DIC modalities were augmented with reflections and masking (illustrated in Fig. \ref{fig:aug}), except for the lattice rotation, which was only augmented with reflections. When considered as augmentation for VAE training, reflection was applied with a probability of 0.5. For masking, each event can be masked with a probability of 0.25 but at least one event per window remained unmasked.

\justify The best plastic deformation intensity VAE has been retrained to integrate data specific to grain boundary sliding which are not yet part of the dataset. For that, the VAE architecture has been retrained for up to 1,000 epochs with a learning rate of $1 \cdot 10^{-3}$. Following the method outlined in Fig. \ref{fig:invar_500}(C), new convolution kernels have been trained based on an initialization with the Glorot initializer \cite{Glorot2010}, with a learning rate factor of 5 (for both the weights and the biases) compared to the unchanged ones (but not frozen).  

\justify The CNN dedicated to properties prediction has been trained using a batch size of 64 for up to 8 epochs (1,000 iterations) and a learning rate of $1 \cdot 10^{-4}$ kept constant during training. 9,216 regions were used to train the architecture while 2,160 regions, from different experimental samples, were utilized for validation. These regions were selected with a maximum overlap of 77\% (7 over 9 pixels) and were subsequently augmented with reflections, Gaussian sampling deriving from the VAE architecture and linear trend between plastic deformation intensity and macroscopic strain (Fig. \ref{fig:aug2}).

\justify All the CNNs utilized in this study have been trained with the Adam updater \cite{Kingma2014}, a gradient decay of $0.9$ and a squared gradient decay of $0.999$. All trainings have been performed using an NVIDIA Geforce RTX 4090.

%\bibliographystyle{unsrtnat}
%\onecolumn

\bibliography{sample} 

\begin{thebibliography}{10}
\urlstyle{rm}
\expandafter\ifx\csname url\endcsname\relax
  \def\url#1{\texttt{#1}}\fi
\expandafter\ifx\csname urlprefix\endcsname\relax\def\urlprefix{URL }\fi
\expandafter\ifx\csname doiprefix\endcsname\relax\def\doiprefix{DOI: }\fi
\providecommand{\bibinfo}[2]{#2}
\providecommand{\eprint}[2][]{\url{#2}}

\bibitem{annurev:/content/journals/10.1146/annurev-matsci-080921-102621}
\bibinfo{author}{Stinville, J.} \emph{et~al.}
\newblock \bibinfo{journal}{\bibinfo{title}{Insights into plastic localization by crystallographic slip from emerging experimental and numerical approaches}}.
\newblock {\emph{\JournalTitle{Annual Review of Materials Research}}} \textbf{\bibinfo{volume}{53}}, \bibinfo{pages}{275--317}, \doiprefix\url{https://doi.org/10.1146/annurev-matsci-080921-102621} (\bibinfo{year}{2023}).

\bibitem{GIANOLA2023101090}
\bibinfo{author}{Gianola, D.~S.} \emph{et~al.}
\newblock \bibinfo{journal}{\bibinfo{title}{Advances and opportunities in high-throughput small-scale mechanical testing}}.
\newblock {\emph{\JournalTitle{Current Opinion in Solid State and Materials Science}}} \textbf{\bibinfo{volume}{27}}, \bibinfo{pages}{101090}, \doiprefix\url{https://doi.org/10.1016/j.cossms.2023.101090} (\bibinfo{year}{2023}).

\bibitem{KAUFMANN2020178}
\bibinfo{author}{Kaufmann, K.} \& \bibinfo{author}{Vecchio, K.~S.}
\newblock \bibinfo{journal}{\bibinfo{title}{Searching for high entropy alloys: A machine learning approach}}.
\newblock {\emph{\JournalTitle{Acta Materialia}}} \textbf{\bibinfo{volume}{198}}, \bibinfo{pages}{178--222}, \doiprefix\url{https://doi.org/10.1016/j.actamat.2020.07.065} (\bibinfo{year}{2020}).

\bibitem{CHENG2025101429}
\bibinfo{author}{Cheng, C.} \& \bibinfo{author}{Zou, Y.}
\newblock \bibinfo{journal}{\bibinfo{title}{Accelerated discovery of nanostructured high-entropy alloys and multicomponent alloys via high-throughput strategies}}.
\newblock {\emph{\JournalTitle{Progress in Materials Science}}} \textbf{\bibinfo{volume}{151}}, \bibinfo{pages}{101429}, \doiprefix\url{https://doi.org/10.1016/j.pmatsci.2025.101429} (\bibinfo{year}{2025}).

\bibitem{Miracle2021131}
\bibinfo{author}{Miracle, D.~B.}, \bibinfo{author}{Li, M.}, \bibinfo{author}{Zhang, Z.}, \bibinfo{author}{Mishra, R.} \& \bibinfo{author}{Flores, K.~M.}
\newblock \bibinfo{journal}{\bibinfo{title}{Emerging capabilities for the high-throughput characterization of structural materials}}.
\newblock {\emph{\JournalTitle{Annual Review of Materials Research}}} \textbf{\bibinfo{volume}{51}}, \bibinfo{pages}{131 – 164}, \doiprefix\url{10.1146/annurev-matsci-080619-022100} (\bibinfo{year}{2021}).

\bibitem{WINIARSKI2021113315}
\bibinfo{author}{Winiarski, B.} \emph{et~al.}
\newblock \bibinfo{journal}{\bibinfo{title}{Correction of artefacts associated with large area ebsd}}.
\newblock {\emph{\JournalTitle{Ultramicroscopy}}} \textbf{\bibinfo{volume}{226}}, \bibinfo{pages}{113315}, \doiprefix\url{https://doi.org/10.1016/j.ultramic.2021.113315} (\bibinfo{year}{2021}).

\bibitem{Black2023}
\bibinfo{author}{Black, R.~L.} \emph{et~al.}
\newblock \bibinfo{journal}{\bibinfo{title}{High-throughput high-resolution digital image correlation measurements by multi-beam sem imaging}}.
\newblock {\emph{\JournalTitle{Experimental Mechanics}}} \textbf{\bibinfo{volume}{63}}, \bibinfo{pages}{939–953}, \doiprefix\url{10.1007/s11340-023-00961-y} (\bibinfo{year}{2023}).

\bibitem{Echlin2021}
\bibinfo{author}{Echlin, M.~P.} \emph{et~al.}
\newblock \bibinfo{journal}{\bibinfo{title}{Recent developments in femtosecond laser-enabled tribeam systems}}.
\newblock {\emph{\JournalTitle{JOM}}} \textbf{\bibinfo{volume}{73}}, \bibinfo{pages}{4258–4269}, \doiprefix\url{10.1007/s11837-021-04919-0} (\bibinfo{year}{2021}).

\bibitem{Stinville2022}
\bibinfo{author}{Stinville, J.} \emph{et~al.}
\newblock \bibinfo{journal}{\bibinfo{title}{On the origins of fatigue strength in crystalline metallic materials}}.
\newblock {\emph{\JournalTitle{Science}}} \textbf{\bibinfo{volume}{377}}, \bibinfo{pages}{1065--1071} (\bibinfo{year}{2022}).

\bibitem{Burnett2019}
\bibinfo{author}{Burnett, T.~L.} \& \bibinfo{author}{Withers, P.~J.}
\newblock \bibinfo{journal}{\bibinfo{title}{Completing the picture through correlative characterization}}.
\newblock {\emph{\JournalTitle{Nature Materials}}} \textbf{\bibinfo{volume}{18}}, \bibinfo{pages}{1041–1049}, \doiprefix\url{10.1038/s41563-019-0402-8} (\bibinfo{year}{2019}).

\bibitem{KAUFMANN2024101192}
\bibinfo{author}{Kaufmann, K.} \& \bibinfo{author}{Vecchio, K.~S.}
\newblock \bibinfo{journal}{\bibinfo{title}{Autonomous materials research and design: Characterization}}.
\newblock {\emph{\JournalTitle{Current Opinion in Solid State and Materials Science}}} \textbf{\bibinfo{volume}{32}}, \bibinfo{pages}{101192}, \doiprefix\url{https://doi.org/10.1016/j.cossms.2024.101192} (\bibinfo{year}{2024}).

\bibitem{Burnett2014}
\bibinfo{author}{Burnett, T.~L.} \emph{et~al.}
\newblock \bibinfo{journal}{\bibinfo{title}{Correlative tomography}}.
\newblock {\emph{\JournalTitle{Scientific Reports}}} \textbf{\bibinfo{volume}{4}}, \doiprefix\url{10.1038/srep04711} (\bibinfo{year}{2014}).

\bibitem{CHARPAGNE2021117037}
\bibinfo{author}{Charpagne, M.} \emph{et~al.}
\newblock \bibinfo{journal}{\bibinfo{title}{Slip localization in inconel 718: A three-dimensional and statistical perspective}}.
\newblock {\emph{\JournalTitle{Acta Materialia}}} \textbf{\bibinfo{volume}{215}}, \bibinfo{pages}{117037}, \doiprefix\url{https://doi.org/10.1016/j.actamat.2021.117037} (\bibinfo{year}{2021}).

\bibitem{STINVILLE2022111891}
\bibinfo{author}{Stinville, J.} \emph{et~al.}
\newblock \bibinfo{journal}{\bibinfo{title}{Observation of bulk plasticity in a polycrystalline titanium alloy by diffraction contrast tomography and topotomography}}.
\newblock {\emph{\JournalTitle{Materials Characterization}}} \textbf{\bibinfo{volume}{188}}, \bibinfo{pages}{111891}, \doiprefix\url{https://doi.org/10.1016/j.matchar.2022.111891} (\bibinfo{year}{2022}).

\bibitem{CHARPAGNE2020110245}
\bibinfo{author}{Charpagne, M.} \emph{et~al.}
\newblock \bibinfo{journal}{\bibinfo{title}{Automated and quantitative analysis of plastic strain localization via multi-modal data recombination}}.
\newblock {\emph{\JournalTitle{Materials Characterization}}} \textbf{\bibinfo{volume}{163}}, \bibinfo{pages}{110245}, \doiprefix\url{https://doi.org/10.1016/j.matchar.2020.110245} (\bibinfo{year}{2020}).

\bibitem{doi:10.1073/pnas.1922210117}
\bibinfo{author}{Lu, L.} \emph{et~al.}
\newblock \bibinfo{journal}{\bibinfo{title}{Extraction of mechanical properties of materials through deep learning from instrumented indentation}}.
\newblock {\emph{\JournalTitle{Proceedings of the National Academy of Sciences}}} \textbf{\bibinfo{volume}{117}}, \bibinfo{pages}{7052--7062}, \doiprefix\url{10.1073/pnas.1922210117} (\bibinfo{year}{2020}).
\newblock \eprint{https://www.pnas.org/doi/pdf/10.1073/pnas.1922210117}.

\bibitem{Thomas2023}
\bibinfo{author}{Thomas, A.} \emph{et~al.}
\newblock \bibinfo{journal}{\bibinfo{title}{Materials fatigue prediction using graph neural networks on microstructure representations}}.
\newblock {\emph{\JournalTitle{Scientific Reports}}} \textbf{\bibinfo{volume}{13}}, \doiprefix\url{10.1038/s41598-023-39400-2} (\bibinfo{year}{2023}).

\bibitem{Raabe2020}
\bibinfo{author}{Raabe, D.} \emph{et~al.}
\newblock \bibinfo{journal}{\bibinfo{title}{Current challenges and opportunities in microstructure-related properties of advanced high-strength steels}}.
\newblock {\emph{\JournalTitle{Metallurgical and Materials Transactions A}}} \textbf{\bibinfo{volume}{51}}, \bibinfo{pages}{5517–5586}, \doiprefix\url{10.1007/s11661-020-05947-2} (\bibinfo{year}{2020}).

\bibitem{GRIFFITHS2021110815}
\bibinfo{author}{Griffiths, S.} \emph{et~al.}
\newblock \bibinfo{journal}{\bibinfo{title}{Influence of hf on the heat treatment response of additively manufactured ni-base superalloy cm247lc}}.
\newblock {\emph{\JournalTitle{Materials Characterization}}} \textbf{\bibinfo{volume}{171}}, \bibinfo{pages}{110815}, \doiprefix\url{https://doi.org/10.1016/j.matchar.2020.110815} (\bibinfo{year}{2021}).

\bibitem{Valle2017}
\bibinfo{author}{Valle, V.} \& \bibinfo{author}{Hedan, S.}
\newblock \bibinfo{journal}{\bibinfo{title}{{Crack Analysis in Mudbricks under Compression Using Specific Development of Stereo-Digital Image Correlation}}}.
\newblock {\emph{\JournalTitle{Experimental Mechanics}}} \doiprefix\url{10.1007/s11340-017-0363-2} (\bibinfo{year}{2017}).

\bibitem{Bourdin2018}
\bibinfo{author}{Bourdin, F.} \emph{et~al.}
\newblock \bibinfo{journal}{\bibinfo{title}{Measurements of plastic localization by heaviside-digital image correlation}}.
\newblock {\emph{\JournalTitle{Acta Materialia}}} \textbf{\bibinfo{volume}{157}}, \bibinfo{pages}{307 -- 325}, \doiprefix\url{https://doi.org/10.1016/j.actamat.2018.07.013} (\bibinfo{year}{2018}).

\bibitem{Texier2024}
\bibinfo{author}{Texier, D.} \emph{et~al.}
\newblock \bibinfo{journal}{\bibinfo{title}{Strain localization in the alloy 718 ni-based superalloy: From room temperature to 650 °c}}.
\newblock {\emph{\JournalTitle{Acta Materialia}}} \textbf{\bibinfo{volume}{268}}, \bibinfo{pages}{119759}, \doiprefix\url{https://doi.org/10.1016/j.actamat.2024.119759} (\bibinfo{year}{2024}).

\bibitem{Anjaria2024}
\bibinfo{author}{Anjaria, D.} \emph{et~al.}
\newblock \bibinfo{journal}{\bibinfo{title}{Plastic deformation delocalization at cryogenic temperatures in a nickel-based superalloy}}.
\newblock {\emph{\JournalTitle{Acta Materialia}}} \textbf{\bibinfo{volume}{276}}, \bibinfo{pages}{120106}, \doiprefix\url{https://doi.org/10.1016/j.actamat.2024.120106} (\bibinfo{year}{2024}).

\bibitem{STINVILLE2020110600}
\bibinfo{author}{Stinville, J.} \emph{et~al.}
\newblock \bibinfo{journal}{\bibinfo{title}{Measurement of elastic and rotation fields during irreversible deformation using heaviside-digital image correlation}}.
\newblock {\emph{\JournalTitle{Materials Characterization}}} \textbf{\bibinfo{volume}{169}}, \bibinfo{pages}{110600}, \doiprefix\url{https://doi.org/10.1016/j.matchar.2020.110600} (\bibinfo{year}{2020}).

\bibitem{He2016}
\bibinfo{author}{He, K.}, \bibinfo{author}{Zhang, X.}, \bibinfo{author}{Ren, S.} \& \bibinfo{author}{Sun, J.}
\newblock \bibinfo{title}{Deep residual learning for image recognition}.
\newblock In \emph{\bibinfo{booktitle}{Proceedings of the IEEE conference on computer vision and pattern recognition}}, \bibinfo{pages}{770--778} (\bibinfo{year}{2016}).

\bibitem{Doersch2016}
\bibinfo{author}{Doersch, C.}
\newblock \bibinfo{journal}{\bibinfo{title}{Tutorial on variational autoencoders}}.
\newblock {\emph{\JournalTitle{arXiv preprint arXiv:1606.05908}}}  (\bibinfo{year}{2016}).

\bibitem{Bean2025}
\bibinfo{author}{Bean, C.} \emph{et~al.}
\newblock \bibinfo{journal}{\bibinfo{title}{Statistical analyses of plastic deformation events via computer vision: Case study of additive manufactured microstructures}}.
\newblock {\emph{\JournalTitle{Materials Characterization}}}  (\bibinfo{year}{2025}).
\newblock \bibinfo{note}{Manuscript submitted for publication}.

\bibitem{Wang2004}
\bibinfo{author}{Wang, Z.}, \bibinfo{author}{Bovik, A.~C.}, \bibinfo{author}{Sheikh, H.~R.} \& \bibinfo{author}{Simoncelli, E.~P.}
\newblock \bibinfo{journal}{\bibinfo{title}{Image quality assessment: from error visibility to structural similarity}}.
\newblock {\emph{\JournalTitle{IEEE transactions on image processing}}} \textbf{\bibinfo{volume}{13}}, \bibinfo{pages}{600--612} (\bibinfo{year}{2004}).

\bibitem{Hore2010}
\bibinfo{author}{Hore, A.} \& \bibinfo{author}{Ziou, D.}
\newblock \bibinfo{title}{Image quality metrics: Psnr vs. ssim}.
\newblock In \emph{\bibinfo{booktitle}{2010 20th international conference on pattern recognition}}, \bibinfo{pages}{2366--2369} (\bibinfo{organization}{IEEE}, \bibinfo{year}{2010}).

\bibitem{Gonzalez2009}
\bibinfo{author}{Gonzalez, R.~C.}
\newblock \emph{\bibinfo{title}{Digital image processing}} (\bibinfo{publisher}{Pearson education india}, \bibinfo{year}{2009}).

\bibitem{Vaswani2017}
\bibinfo{author}{Vaswani, A.}
\newblock \bibinfo{journal}{\bibinfo{title}{Attention is all you need}}.
\newblock {\emph{\JournalTitle{Advances in Neural Information Processing Systems}}}  (\bibinfo{year}{2017}).

\bibitem{Kingma2013}
\bibinfo{author}{Kingma, D.~P.}
\newblock \bibinfo{journal}{\bibinfo{title}{Auto-encoding variational bayes}}.
\newblock {\emph{\JournalTitle{arXiv preprint arXiv:1312.6114}}}  (\bibinfo{year}{2013}).

\bibitem{bean2025acceleratedfatiguestrengthprediction}
\bibinfo{author}{Bean, C.} \emph{et~al.}
\newblock \bibinfo{title}{Accelerated fatigue strength prediction via additive manufactured functionally graded materials and high-throughput plasticity quantification} (\bibinfo{year}{2025}).
\newblock \eprint{2502.13159}.

\bibitem{Mughrabi2009}
\bibinfo{author}{Mughrabi, H.}
\newblock \bibinfo{journal}{\bibinfo{title}{Cyclic slip irreversibilities and the evolution of fatigue damage}}.
\newblock {\emph{\JournalTitle{Metallurgical and Materials Transactions A}}} \textbf{\bibinfo{volume}{40}}, \bibinfo{pages}{1257--1279}, \doiprefix\url{10.1007/s11661-009-9839-8} (\bibinfo{year}{2009}).

\bibitem{STINVILLE2020172}
\bibinfo{author}{Stinville, J.} \emph{et~al.}
\newblock \bibinfo{journal}{\bibinfo{title}{Direct measurements of slip irreversibility in a nickel-based superalloy using high resolution digital image correlation}}.
\newblock {\emph{\JournalTitle{Acta Materialia}}} \textbf{\bibinfo{volume}{186}}, \bibinfo{pages}{172--189}, \doiprefix\url{https://doi.org/10.1016/j.actamat.2019.12.009} (\bibinfo{year}{2020}).

\bibitem{HULL2011171}
\bibinfo{author}{Hull, D.} \& \bibinfo{author}{Bacon, D.}
\newblock \bibinfo{title}{Chapter 9 - dislocation arrays and crystal boundaries}.
\newblock In \bibinfo{editor}{Hull, D.} \& \bibinfo{editor}{Bacon, D.} (eds.) \emph{\bibinfo{booktitle}{Introduction to Dislocations (Fifth Edition)}}, \bibinfo{pages}{171--204}, \doiprefix\url{https://doi.org/10.1016/B978-0-08-096672-4.00009-8} (\bibinfo{publisher}{Butterworth-Heinemann}, \bibinfo{address}{Oxford}, \bibinfo{year}{2011}), \bibinfo{edition}{fifth edition} edn.

\bibitem{Andani2020}
\bibinfo{author}{Andani, M.~T.} \emph{et~al.}
\newblock \bibinfo{journal}{\bibinfo{title}{A quantitative study of stress fields ahead of a slip band blocked by a grain boundary in unalloyed magnesium}}.
\newblock {\emph{\JournalTitle{Scientific Reports}}} \textbf{\bibinfo{volume}{10}}, \doiprefix\url{10.1038/s41598-020-59684-y} (\bibinfo{year}{2020}).

\bibitem{ANDANI2022117613}
\bibinfo{author}{Andani, M.~T.}, \bibinfo{author}{Lakshmanan, A.}, \bibinfo{author}{Sundararaghavan, V.}, \bibinfo{author}{Allison, J.} \& \bibinfo{author}{Misra, A.}
\newblock \bibinfo{journal}{\bibinfo{title}{Estimation of micro-hall-petch coefficients for prismatic slip system in mg-4al as a function of grain boundary parameters}}.
\newblock {\emph{\JournalTitle{Acta Materialia}}} \textbf{\bibinfo{volume}{226}}, \bibinfo{pages}{117613}, \doiprefix\url{https://doi.org/10.1016/j.actamat.2021.117613} (\bibinfo{year}{2022}).

\bibitem{doi:https://doi.org/10.1002/9781119296126.ch173}
\bibinfo{author}{Guo, Y.}, \bibinfo{author}{Britton, T.~B.} \& \bibinfo{author}{Wilkinson, A.~J.}
\newblock \emph{\bibinfo{title}{Stress Concentrations, Slip Bands and Grain Boundaries In Commercially Pure Titanium}}, chap. \bibinfo{chapter}{173}, \bibinfo{pages}{1017--1021} (\bibinfo{publisher}{John Wiley And Sons, Ltd}, \bibinfo{year}{2016}).
\newblock \eprint{https://onlinelibrary.wiley.com/doi/pdf/10.1002/9781119296126.ch173}.

\bibitem{doi:10.1126/sciadv.abo5735}
\bibinfo{author}{Edwards, T. E.~J.}, \bibinfo{author}{Maeder, X.}, \bibinfo{author}{Ast, J.}, \bibinfo{author}{Berger, L.} \& \bibinfo{author}{Michler, J.}
\newblock \bibinfo{journal}{\bibinfo{title}{Mapping pure plastic strains against locally applied stress: Revealing toughening plasticity}}.
\newblock {\emph{\JournalTitle{Science Advances}}} \textbf{\bibinfo{volume}{8}}, \bibinfo{pages}{eabo5735}, \doiprefix\url{10.1126/sciadv.abo5735} (\bibinfo{year}{2022}).
\newblock \eprint{https://www.science.org/doi/pdf/10.1126/sciadv.abo5735}.

\bibitem{CHAKRAVARTHY2010625}
\bibinfo{author}{Chakravarthy, S.~S.} \& \bibinfo{author}{Curtin, W.}
\newblock \bibinfo{journal}{\bibinfo{title}{Effect of source and obstacle strengths on yield stress: A discrete dislocation study}}.
\newblock {\emph{\JournalTitle{Journal of the Mechanics and Physics of Solids}}} \textbf{\bibinfo{volume}{58}}, \bibinfo{pages}{625 -- 635}, \doiprefix\url{https://doi.org/10.1016/j.jmps.2010.03.004} (\bibinfo{year}{2010}).

\bibitem{CLEVERINGA1999837}
\bibinfo{author}{Cleveringa, H.}, \bibinfo{author}{{Van der Giessen}, E.} \& \bibinfo{author}{Needleman, A.}
\newblock \bibinfo{journal}{\bibinfo{title}{A discrete dislocation analysis of bending}}.
\newblock {\emph{\JournalTitle{International Journal of Plasticity}}} \textbf{\bibinfo{volume}{15}}, \bibinfo{pages}{837--868}, \doiprefix\url{https://doi.org/10.1016/S0749-6419(99)00013-3} (\bibinfo{year}{1999}).

\bibitem{refId0}
\bibinfo{author}{{Cleveringa, H.}}, \bibinfo{author}{{Van der Giessen, E.}} \& \bibinfo{author}{{Needleman, A.}}
\newblock \bibinfo{journal}{\bibinfo{title}{Discrete dislocation simulations and size dependent hardening in single slip}}.
\newblock {\emph{\JournalTitle{J. Phys. IV France}}} \textbf{\bibinfo{volume}{08}}, \bibinfo{pages}{Pr4--83--Pr4--92}, \doiprefix\url{10.1051/jp4:1998410} (\bibinfo{year}{1998}).

\bibitem{BITTENCOURT2018169}
\bibinfo{author}{Bittencourt, E.}
\newblock \bibinfo{journal}{\bibinfo{title}{On the effects of hardening models and lattice rotations in strain gradient crystal plasticity simulations}}.
\newblock {\emph{\JournalTitle{International Journal of Plasticity}}} \textbf{\bibinfo{volume}{108}}, \bibinfo{pages}{169--185}, \doiprefix\url{https://doi.org/10.1016/j.ijplas.2018.05.004} (\bibinfo{year}{2018}).

\bibitem{Vermeij2023}
\bibinfo{author}{Vermeij, T.}, \bibinfo{author}{Peerlings, R.}, \bibinfo{author}{Geers, M.} \& \bibinfo{author}{Hoefnagels, J.}
\newblock \bibinfo{journal}{\bibinfo{title}{Automated identification of slip system activity fields from digital image correlation data}}.
\newblock {\emph{\JournalTitle{Acta Materialia}}} \textbf{\bibinfo{volume}{243}}, \bibinfo{pages}{118502}, \doiprefix\url{https://doi.org/10.1016/j.actamat.2022.118502} (\bibinfo{year}{2023}).

\bibitem{Charpagne2020}
\bibinfo{author}{Charpagne, M.} \emph{et~al.}
\newblock \bibinfo{journal}{\bibinfo{title}{Automated and quantitative analysis of plastic strain localization via multi-modal data recombination}}.
\newblock {\emph{\JournalTitle{Materials Characterization}}} \textbf{\bibinfo{volume}{163}}, \bibinfo{pages}{110245}, \doiprefix\url{https://doi.org/10.1016/j.matchar.2020.110245} (\bibinfo{year}{2020}).

\bibitem{Hu2023}
\bibinfo{author}{Hu, H.}, \bibinfo{author}{Briffod, F.}, \bibinfo{author}{Yin, W.}, \bibinfo{author}{Shiraiwa, T.} \& \bibinfo{author}{Enoki, M.}
\newblock \bibinfo{journal}{\bibinfo{title}{Quantitative investigation of slip band activities in a bimodal titanium alloy under pure fatigue and dwell-fatigue loadings}}.
\newblock {\emph{\JournalTitle{International Journal of Fatigue}}} \textbf{\bibinfo{volume}{182}}, \bibinfo{pages}{108203}, \doiprefix\url{https://doi.org/10.1016/j.ijfatigue.2024.108203} (\bibinfo{year}{2024}).

\bibitem{Ni2024}
\bibinfo{author}{Ni, R.} \emph{et~al.}
\newblock \bibinfo{journal}{\bibinfo{title}{Automated analysis framework of strain partitioning and deformation mechanisms via multimodal fusion and computer vision}}.
\newblock {\emph{\JournalTitle{International Journal of Plasticity}}} \textbf{\bibinfo{volume}{182}}, \bibinfo{pages}{104119}, \doiprefix\url{https://doi.org/10.1016/j.ijplas.2024.104119} (\bibinfo{year}{2024}).

\bibitem{Calvat2025}
\bibinfo{author}{Calvat, M.} \emph{et~al.}
\newblock \bibinfo{journal}{\bibinfo{title}{Learning metal microstructural heterogeneity through spatial mapping of diffraction latent space features}}.
\newblock {\emph{\JournalTitle{arXiv preprint arXiv:2501.18064}}}  (\bibinfo{year}{2025}).

\bibitem{Biswas2023}
\bibinfo{author}{Biswas, A.}, \bibinfo{author}{Ziatdinov, M.} \& \bibinfo{author}{Kalinin, S.~V.}
\newblock \bibinfo{journal}{\bibinfo{title}{Combining variational autoencoders and physical bias for improved microscopy data analysis}}.
\newblock {\emph{\JournalTitle{Machine Learning: Science and Technology}}} \textbf{\bibinfo{volume}{4}}, \bibinfo{pages}{045004} (\bibinfo{year}{2023}).

\bibitem{Valleti2024}
\bibinfo{author}{Valleti, M.}, \bibinfo{author}{Ziatdinov, M.}, \bibinfo{author}{Liu, Y.} \& \bibinfo{author}{Kalinin, S.~V.}
\newblock \bibinfo{journal}{\bibinfo{title}{Physics and chemistry from parsimonious representations: image analysis via invariant variational autoencoders}}.
\newblock {\emph{\JournalTitle{npj Computational Materials}}} \textbf{\bibinfo{volume}{10}}, \bibinfo{pages}{183} (\bibinfo{year}{2024}).

\bibitem{Goodfellow2016}
\bibinfo{author}{Goodfellow, I.}, \bibinfo{author}{Bengio, Y.} \& \bibinfo{author}{Courville, A.}
\newblock \emph{\bibinfo{title}{Deep Learning}} (\bibinfo{publisher}{MIT Press}, \bibinfo{year}{2016}).
\newblock \bibinfo{note}{\url{http://www.deeplearningbook.org}}.

\bibitem{MANGAL2018122}
\bibinfo{author}{Mangal, A.} \& \bibinfo{author}{Holm, E.~A.}
\newblock \bibinfo{journal}{\bibinfo{title}{Applied machine learning to predict stress hotspots i: Face centered cubic materials}}.
\newblock {\emph{\JournalTitle{International Journal of Plasticity}}} \textbf{\bibinfo{volume}{111}}, \bibinfo{pages}{122--134}, \doiprefix\url{https://doi.org/10.1016/j.ijplas.2018.07.013} (\bibinfo{year}{2018}).

\bibitem{MANGAL20191}
\bibinfo{author}{Mangal, A.} \& \bibinfo{author}{Holm, E.~A.}
\newblock \bibinfo{journal}{\bibinfo{title}{Applied machine learning to predict stress hotspots ii: Hexagonal close packed materials}}.
\newblock {\emph{\JournalTitle{International Journal of Plasticity}}} \textbf{\bibinfo{volume}{114}}, \bibinfo{pages}{1--14}, \doiprefix\url{https://doi.org/10.1016/j.ijplas.2018.08.003} (\bibinfo{year}{2019}).

\bibitem{Dai2021}
\bibinfo{author}{Dai, M.}, \bibinfo{author}{Demirel, M.~F.}, \bibinfo{author}{Liang, Y.} \& \bibinfo{author}{Hu, J.-M.}
\newblock \bibinfo{journal}{\bibinfo{title}{Graph neural networks for an accurate and interpretable prediction of the properties of polycrystalline materials}}.
\newblock {\emph{\JournalTitle{npj Computational Materials}}} \textbf{\bibinfo{volume}{7}}, \doiprefix\url{10.1038/s41524-021-00574-w} (\bibinfo{year}{2021}).

\bibitem{SU201940}
\bibinfo{author}{Su, J.}, \bibinfo{author}{Raabe, D.} \& \bibinfo{author}{Li, Z.}
\newblock \bibinfo{journal}{\bibinfo{title}{Hierarchical microstructure design to tune the mechanical behavior of an interstitial trip-twip high-entropy alloy}}.
\newblock {\emph{\JournalTitle{Acta Materialia}}} \textbf{\bibinfo{volume}{163}}, \bibinfo{pages}{40--54}, \doiprefix\url{https://doi.org/10.1016/j.actamat.2018.10.017} (\bibinfo{year}{2019}).

\bibitem{Mcinnes2018}
\bibinfo{author}{McInnes, L.}, \bibinfo{author}{Healy, J.} \& \bibinfo{author}{Melville, J.}
\newblock \bibinfo{journal}{\bibinfo{title}{Umap: Uniform manifold approximation and projection for dimension reduction}}.
\newblock {\emph{\JournalTitle{arXiv preprint arXiv:1802.03426}}}  (\bibinfo{year}{2018}).

\bibitem{Weiss2016}
\bibinfo{author}{Weiss, K.}, \bibinfo{author}{Khoshgoftaar, T.~M.} \& \bibinfo{author}{Wang, D.}
\newblock \bibinfo{journal}{\bibinfo{title}{A survey of transfer learning}}.
\newblock {\emph{\JournalTitle{Journal of Big data}}} \textbf{\bibinfo{volume}{3}}, \bibinfo{pages}{1--40} (\bibinfo{year}{2016}).

\bibitem{Kammers_2013a}
\bibinfo{author}{Kammers, A.} \& \bibinfo{author}{Daly, S.}
\newblock \bibinfo{journal}{\bibinfo{title}{Self-assembled nanoparticle surface patterning for improved digital image correlation in a scanning electron microscope}}.
\newblock {\emph{\JournalTitle{Experimental Mechanics}}} \textbf{\bibinfo{volume}{53}}, \bibinfo{pages}{1333--1341}, \doiprefix\url{10.1007/s11340-013-9734-5} (\bibinfo{year}{2013}).

\bibitem{Montgomery2019}
\bibinfo{author}{Montgomery, C.}, \bibinfo{author}{Koohbor, B.} \& \bibinfo{author}{Sottos, N.}
\newblock \bibinfo{journal}{\bibinfo{title}{A robust patterning technique for electron microscopy-based digital image correlation at sub-micron resolutions}}.
\newblock {\emph{\JournalTitle{Experimental Mechanics}}} \textbf{\bibinfo{volume}{59}}, \bibinfo{pages}{1063--1073}, \doiprefix\url{10.1007/s11340-019-00487-2} (\bibinfo{year}{2019}).

\bibitem{Kammers_2013b}
\bibinfo{author}{Kammers, A.} \& \bibinfo{author}{Daly, S.}
\newblock \bibinfo{journal}{\bibinfo{title}{Digital image correlation under scanning electron microscopy: Methodology and validation}}.
\newblock {\emph{\JournalTitle{Experimental Mechanics}}} \textbf{\bibinfo{volume}{53}}, \bibinfo{pages}{1743--1761}, \doiprefix\url{10.1007/s11340-013-9782-x} (\bibinfo{year}{2013}).

\bibitem{Stinville_2015a}
\bibinfo{author}{Stinville, J.} \emph{et~al.}
\newblock \bibinfo{journal}{\bibinfo{title}{Sub-grain scale digital image correlation by electron microscopy for polycrystalline materials during elastic and plastic deformation}}.
\newblock {\emph{\JournalTitle{Experimental Mechanics}}} \bibinfo{pages}{1--20}, \doiprefix\url{10.1007/s11340-015-0083-4} (\bibinfo{year}{2015}).

\bibitem{Mello2017}
\bibinfo{author}{Mello, A.~W.} \emph{et~al.}
\newblock \bibinfo{journal}{\bibinfo{title}{Distortion correction protocol for digital image correlation after scanning electron microscopy: Emphasis on long duration and ex-situ experiments}}.
\newblock {\emph{\JournalTitle{Experimental Mechanics}}} \doiprefix\url{10.1007/s11340-017-0303-1} (\bibinfo{year}{2017}).

\bibitem{Besson2009}
\bibinfo{author}{Besson, J.}, \bibinfo{author}{Cailletaud, G.}, \bibinfo{author}{Chaboche, J.-L.} \& \bibinfo{author}{Forest, S.}
\newblock \emph{\bibinfo{title}{Non-linear mechanics of materials}}, vol. \bibinfo{volume}{167} (\bibinfo{publisher}{Springer Science \& Business Media}, \bibinfo{year}{2009}).

\bibitem{BOURDIN2018307}
\bibinfo{author}{Bourdin, F.} \emph{et~al.}
\newblock \bibinfo{journal}{\bibinfo{title}{Measurements of plastic localization by heaviside-digital image correlation}}.
\newblock {\emph{\JournalTitle{Acta Materialia}}} \textbf{\bibinfo{volume}{157}}, \bibinfo{pages}{307 -- 325}, \doiprefix\url{https://doi.org/10.1016/j.actamat.2018.07.013} (\bibinfo{year}{2018}).

\bibitem{Lu2019}
\bibinfo{author}{Lu, L.}, \bibinfo{author}{Shin, Y.}, \bibinfo{author}{Su, Y.} \& \bibinfo{author}{Karniadakis, G.~E.}
\newblock \bibinfo{journal}{\bibinfo{title}{Dying relu and initialization: Theory and numerical examples}}.
\newblock {\emph{\JournalTitle{arXiv preprint arXiv:1903.06733}}}  (\bibinfo{year}{2019}).

\bibitem{Glorot2010}
\bibinfo{author}{Glorot, X.} \& \bibinfo{author}{Bengio, Y.}
\newblock \bibinfo{title}{Understanding the difficulty of training deep feedforward neural networks}.
\newblock In \emph{\bibinfo{booktitle}{Proceedings of the thirteenth international conference on artificial intelligence and statistics}}, \bibinfo{pages}{249--256} (\bibinfo{organization}{JMLR Workshop and Conference Proceedings}, \bibinfo{year}{2010}).

\bibitem{Kingma2014}
\bibinfo{author}{Kingma, D.~P.}
\newblock \bibinfo{journal}{\bibinfo{title}{Adam: A method for stochastic optimization}}.
\newblock {\emph{\JournalTitle{arXiv preprint arXiv:1412.6980}}}  (\bibinfo{year}{2014}).

\end{thebibliography}

\noindent\textbf{Acknowledgments} \\

\justify M.C., C.B., H.W., K.V. and J.C.S. are grateful for financial support from the the Defense Advanced Research Projects Agency (DARPA - HR001124C0394). This work was carried out in the Materials Research Laboratory Central Research Facilities, University of Illinois, and at the Advanced Materials Testing and Evaluation Laboratory, University of Illinois. Carpenter Technology is acknowledged for providing the W\_RX\_In718 material. Morad Behandish and Adrian Lew are acknowledged for their support and leadership. Acknowledgment is extended to Valery Valle for providing the Heaviside-DIC code. \\

\noindent\textbf{CRediT authorship contribution statement} \\

\justify \textbf{M.C.}: Conceptualization, Data curation, Formal analysis, Investigation, Methodology, Writing – original draft, Writing – review \& editing. \textbf{C.B.}: Conceptualization, Data curation, Formal analysis, Investigation, Methodology, Writing – original draft, Writing – review \& editing. \textbf{D.A.}: Data curation, Verification, Writing – review \& editing. \textbf{H.W.}: Resources. \textbf{K.V.}: Writing – review \& editing. \textbf{J.C.S.}: Conceptualization, Funding acquisition, Methodology, Project administration, Resources, Supervision, Writing – original draft, Writing – review \& editing. \\

\noindent\textbf{Competing interests} \\

\justify The authors declare that they have no known competing financial interests or personal relationships that could have appeared to influence the work reported in this paper. \\

\noindent\textbf{Supplementary materials} \\

\begin{table}[h!]
    \centering
    \setlength{\tabcolsep}{6pt}
    \caption{Performance of reconstruction of the different HR-DIC modalities with all considered augmentations for both longitudinal strain and plastic deformation direction, but only reflections regarding the lattice rotation.}
    \label{tab:perfo2}
    \begin{tabular}{|c||c|c|c|c|c|c|}
        \hline
        Alloy & \multicolumn{2}{c|}{Longitudinal strain} & \multicolumn{2}{c|}{Plastic deformation direction} & \multicolumn{2}{c|}{Lattice rotation} \\
        Denomination & SSIM & PSNR & SSIM & PSNR & SSIM & PSNR \\
        & 1\textsuperscript{st} -- Avg. & 1\textsuperscript{st} -- Avg. & 1\textsuperscript{st} -- Avg. & 1\textsuperscript{st} -- Avg. & 1\textsuperscript{st} -- Avg. & 1\textsuperscript{st} -- Avg. \\
        \hline
        W\_Invar & 0.392 -- 0.296 & 24.60 -- 24.22 & 0.0002 -- 0.0174 & 12.59 -- 13.66 & 0.954 -- 0.874 & 24.46 -- 18.60 \\
        \hline
        W\_Nickel600 & 0.300 -- 0.253 & 22.15 -- 22.12 & 0.0004 -- 0.0012 & 12.74 -- 13.00 & 0.916 -- 0.856 & 23.64 -- 24.60 \\
        \hline
        W\_Steel330 & 0.276 -- 0.258 & 22.43 -- 22.57 & 0.0006 -- 0.0010 & 12.84 -- 13.16 & 0.919 -- 0.891 & 30.222 -- 31.11 \\
        \hline
        W\_In718 & 0.233 -- 0.242 & 21.95 -- 22.52 & 0.0001 -- 0.0006 & 12.64 -- 12.77 & 0.942 -- 0.880 & 26.55 -- 27.37 \\
        \hline
        W\_Cu77NiSn & 0.223 -- 0.225 & 23.58 -- 23.90 & 0.0006 -- 0.0025 & 12.83 -- 13.26 & 0.811 -- 0.822 & 16.80 -- 18.58 \\
        \hline
        W\_RX\_In718 & 0.205 -- 0.237 & 27.52 -- 26.18 & 0.0015 -- 0.0036 & 12.76 -- 13.14 & 0.864 -- 0.824 & 10.28 -- 19.23 \\
        \hline
        AM\_AB\_In718 & 0.198 -- 0.253 & 23.18 -- 23.42 & 0.0012 -- 0.0027 & 12.80 -- 13.19 & 0.937 -- 0.803 & 25.37 -- 22.50 \\
        \hline
        AM\_AB\_Co76A & 0.237 -- 0.255 & 23.40 -- 24.07 & 0.0011 -- 0.0023  & 12.81 -- 13.03 & 0.907 -- 0.794 & 14.22 -- 17.48 \\
        \hline
        AM\_AB\_Steel316 & 0.292 -- 0.313 & 26.14 -- 26.12 & 0.0014 -- 0.0024  & 12.69 -- 12.93 & 0.869 -- 0.791 & 10.84 -- 15.61 \\
        \hline
    \end{tabular}
\end{table}

\begin{table}[h!]
    \centering
    \setlength{\tabcolsep}{6pt}
    \caption{Relative error regarding the mechanical properties prediction for the different materials. Extreme low and high values are given and have been estimated from the additional samples.}
    \label{tab:acc}
    \begin{tabular}{|c||c|c|c|c|c|c|}
        \hline
        Alloy & \multicolumn{6}{c|}{Relative error (\%)} \\
        Denomination & YS & Q & b & UTS & El & FS \\
        \hline
        W\_Invar & 26.4 -- 34.4 & 15.0 -- 18.7 & 9.9 -- 16.0 & 35.3 -- 41.2 & 4.6 -- 7.6 & 31.6 -- 35.3 \\
        \hline
        W\_Nickel600 & 13.2 -- 24.0 & 19.4 -- 29.5 & 6.6 -- 13.0 & 1.4 -- 5.7 & 0.0 -- 2.3 & 4.5 -- 6.7 \\
        \hline
        W\_Steel330 & 2.3 -- 7.8 & 7.6 -- 12.8 & 16.4 -- 19.1 & 3.9 -- 9.8 & 2.9 -- 5.4 & 0.0 -- 2.1 \\
        \hline
        W\_In718 & 15.3 -- 20.2 & 48.3 -- 51.2 & 15.0 -- 19.3 & 22.0 -- 24.8 & 17.4 -- 20.3 & 16.3 -- 17.4 \\
        \hline
        W\_Cu77NiSn & 8.5 -- 11.9 & 57.8 -- 64.6 & 2.6 -- 6.2 & 20.0 -- 23.0 & 27.7 -- 34.1 & 20.8 -- 23.9 \\
        \hline
        W\_RX\_In718 & 30.1 -- 32.4 & 20.0 -- 24.6 & 37.8 -- 41.3 & 22.8 -- 25.2 & 9.1 -- 14.2 & 8.7 -- 11.5 \\
        \hline
        AM\_AB\_In718 & 9.7 -- 13.0 & 10.8 -- 15.4 & 1.9 -- 4.1 & 4.8 -- 7.4 & 3.7 -- 10.6 & 10.8 -- 14.6  \\
        \hline
        AM\_AB\_Co76A & 5.4 -- 9.5 & 36.8 -- 41.9 & 7.9 -- 10.6 & 17.3 -- 20.5 & 14.2 -- 25.9 & 18.0 -- 20.8 \\
        \hline
        AM\_AB\_Steel316 & 14.9 -- 20.7 & 9.8 -- 16.5 & 13.1 -- 17.5 & 5.3 -- 9.0 & 18.2 -- 22.5 & 0.0 -- 1.3 \\
        \hline
    \end{tabular}
\end{table}

\end{document}